\documentclass[preprint,secnumarabic,amssymb,showpacs,showkeys,nobibnotes, aps, prd,nofootinbib]{revtex4-1}

\usepackage{graphicx}
\usepackage{amsmath}
\begin{document}
\title{NULL GEODESICS IN A MAGNETICALLY CHARGED STRINGY BLACK HOLE SPACETIME}
\author{Ravi Shankar Kuniyal}
\email{ravikuniyal09@gmail.com}
\affiliation{Department of Physics, Gurukula Kangri Vishwavidyalaya,\\
Haridwar 249 404, Uttarakhand, India.}
\author{Rashmi Uniyal}
\email{rashmiuniyal001@gmail.com}
\affiliation{Department of Physics, Gurukula Kangri Vishwavidyalaya,\\
Haridwar 249 404, Uttarakhand, India.}
\author{Hemwati Nandan}
\email{hnandan@iucca.ernet.in}
\affiliation{Department of Physics, Gurukula Kangri Vishwavidyalaya,\\
Haridwar 249 404, Uttarakhand, India.}
\author{K. D. Purohit}
\email{kdpurohit@gmail.com}
\affiliation{Department of Physics, HNB Garhwal University,\\
Srinagar Garhwal 246 174, Uttarakhand, India.}
\begin{abstract}
We study the geodesic motion of massless test particles in the background of a magnetic charged black hole spacetime in four dimensions in dilaton-Maxwell gravity. The behaviour of effective potential in view of the different values of black hole parameters is analysed in the equatorial plane. The possible orbits for null geodesics are also discussed in detail in view of the different values of the impact parameter. We have also calculated the frequency shift of photons in this spacetime. The results obtained are then compared with those for the electrically charged stringy black hole spacetime and the Schwarzschild black hole spacetime. It is observed that there exists no stable circular orbit outside the event horizon for massless test particles.
\end{abstract}
\maketitle
\tableofcontents
\section{Introduction}
\noindent Black holes (BHs) arising in Einstein's General Relativity (GR) are one of the most fascinating objects in the universe \cite{har03}. GR has developed into an essential tool in modern astrophysics which provides the foundation for the current understanding of BHs. The most general spherically symmetric, vacuum solution of the Einstein field equations in GR is the well known Schwarzschild black hole (SBH) spacetime \cite{har03,cha83}. Further, a static solution to the Einstein-Maxwell field equations, which corresponds to the gravitational field of a charged, non-rotating, spherically symmetric body is the Reissner-Nordstr$\ddot{o}$m spacetime \cite{har03,cha83}. The study of charged BHs may helpful to understand more general spacetimes with sources immersed into spherically symmetric matter backgrounds \cite{pugl11}. Next, Kerr black hole spacetime is a rotating generalization of the SBH spacetime and the Kerr-Newman BH spacetime which is a solution of Einstein-Maxwell equations in GR, describes the spacetime geometry in the region surrounding a charged rotating mass \cite{sch85, eva13}.\\
In addition to the abovementioned BH spacetimes those arise in GR, there are BH spacetime solutions in various alternative theories of gravity viz. scalar-tensor theory \cite{fuj03}, string theory \cite{ton09} and loop quantum gravity \cite{thi03}  etc.. In particular, the string theory is an important candidate for unifying gravity with other three fundamental forces in nature and therefore BH spacetimes emerging in the string theory are of much interest as they may provide much deeper insight into the various properties of BH spacetimes than the BH spacetimes in GR \cite{kar99, sen92}.
Lower dimensional charged dilaton BH has a connection to the lower dimensional string theory and hence it might act as a tool to study lower dimensional string theory \cite{mig93,sen92, gib88, kar99}.\\
The study of geodesic structure of massless particles in a given BH spacetime is one of the important ways to understand the gravitational field around a BH spactime. 
The geodesic motion around various BH spacetimes in diverse contexts (for timelike as well as null geodesics), both in GR and string theory, are studied widely time and again \cite{dab97, fern12, fer12, fer14, nor05, pug11, hio08, bha03, pra11, stu14, sch09, kol03, uni14, ras15, eva10, mak94, fer03, eva08, nan08, anv08, das09, das12, gho10, rav15}.\\
The primary aim of the present work is to study the null geodesics in a four dimensional spherically symmetric BH spacetime with magnetic charge, obtained from the low energy effective action in string theory \cite{ghs91, gth92}.
The circular timelike and null geodesics in equatorial plane are discussed with some new results recently \cite{pra12}. Also the unstable circular orbit of photons for electrically charged black hole (ECBH) in string frame is already discussed \cite{fern12}. Other than the circular orbit, we are also interested in all other kind of possible orbits depending upon the value of the impact parameter.\\
This paper is organised as follows. A brief introduction to the spacetime considered is presented in the next section. We solve the geodesic equations for null geodesics for magnetically charged black hole (MCBH) spacetime and discuss the effective potential for BH parameters in section III. We also analyze the photon orbits and then compare the results with the ECBH and SBH. As one of the measurable effect of GR, we have also calculated the frequency shift for the MCBH. Finally, in the last section, we have summarized our results with future possibilities.
\section{Magnetically charged black hole spacetime in string theory}
\noindent
In 3 + 1 dimensions, Garfinkle, Horowitz and Strominger \cite{gth92} have obtained asymptotically flat solutions representing the BHs in dilaton-Maxwell gravity.
Such solutions as a specific feature, represent the electric and dual magnetic BHs \cite{gth92}.
The metric for the MCBH can be deduced from the ECBH by an electromagnetic duality transformation and is given as,
\begin{equation}
ds^2= -\frac{(1-\frac{2 m}{r})}{(1-\frac{Q^2}{m r})}dt^2+\frac{1}{(1-\frac{2 m}{r})(1-\frac{Q^2}{m r})}dr^2+r^2 d\Omega^2 ,
\label{eq:magnetic}
\end{equation}
where $Q$ is the magnetic charge and $m$ is the mass of the BH while $d\Omega^2=d\theta^2+\sin^2\theta d\phi^2$ represents a metric on a two dimensional unit sphere.
Event horizon of the BH is situated at $r$ = 2$m$ and the singularity occurs at $r=\frac{Q^2}{m}$. The metric for the (corresponding) ECBH is given as,
\begin{equation}
ds^2= -\frac{(1-\frac{2 m}{r})}{(1+\frac{2 m \sinh^2(\alpha)}{r})^2}dt^2 + \frac{1}{(1-\frac{2 m}{r})} dr^2 + r^2 d\Omega^2 ,
\label{eq:electric}
\end{equation}
where $m$ is the mass of BH and $\alpha$ is a parameter related to the electric charge.
The position of event horizon is at $r=2 m$ and curvature singularity occurs at $r=0$.
One important feature to note is that the string coupling is strong near singularity for the MCBH while it is weak for the ECBH \cite{gth92}.
The spacetime geometry of these dual BH solutions are causally similar to Schwarzschild geometry.
In the respective limits of $Q = 0$ and $\alpha =0 $, the abovementioned spacetimes reduce to the SBH spacetime.
\section{Null geodesics for magnetically charged black hole}
\noindent In this section, we study the radial and non radial null geodesics alongwith the possible orbit structure for spacetime given by eq.(\ref{eq:magnetic}) in the equatorial plane (i.e. $\theta=\pi/2$).
The geodesic equations and constraints for null geodesics are given as, $\ddot{x^a}+\Gamma^{a}_{bc}\dot{x}^{b}\dot{x}^{c}=0 $ and  $g_{{a}{b}}\dot{x}^a \dot{x}^b=0$ respectively \cite{das09} (where dot denotes the differentiation with respect to an affine parameter).
Now using the geodesic equations their first integrals can be expressed as,
\begin{eqnarray}
\dot{t}=\frac{E (m r - Q^2)}{m (r - 2 m)},\\
\label{eqn:first_int_t}
\dot{\phi}=\frac{L}{r^2}.
\label{eqn:first_intg_M}
\end{eqnarray}
The constraint on null geodesics for the line element given by eq.(\ref{eq:magnetic}) reads as,
\begin{equation}
-\frac{(1 - \frac{2 m}{r})}{(1 - \frac{Q^2}{m r})}\dot{t}^2 + \left[(1 - \frac{2 m}{r})(1 - \frac{Q^2}{m r})\right]^{- 1}\dot{r}^2 + r^2(\dot{\theta}^2 + \sin^2\theta \dot{\phi}^2) = 0 .
\label{eqn:cons_M}
\end{equation}
Further using equation (\ref{eqn:first_intg_M}) and equation (\ref{eqn:cons_M}), the radial velocity in the equatorial plane is given as,
\begin{equation}
\dot{r}^2=L^2\left[\frac{(m r-Q^2)^2}{D^2 m^2 r^2} - \frac{(r - 2 m)(m r - Q^2)}{m r^4}\right] ,
\label{eqn:vel_M}
\end{equation}
where $D = L/E$ is the impact parameter \cite{cha83,har03}. Hence the effective potential for null geodesics is defined as,
\begin{equation}
V(r) = L^2 \left[\frac{(r - 2 m)(m r - Q^2)}{m r^4}-\frac{(m r - Q^2)^2}{D^2 m^2 r^2}\right] .
\label{eqn:eff_potn_M}
\end{equation}
In case of null geodesics, the impact parameter plays a similar role as that of the energy of the incoming test particle in case of timelike geodesics \cite{har03}.
Hence, the types of orbit for massless test particles (i.e. photons) depend on their impact parameter.
\subsection{Radial null geodesics}
\begin{figure}[h]
\centerline{\includegraphics[width=6.5cm,height=6cm]{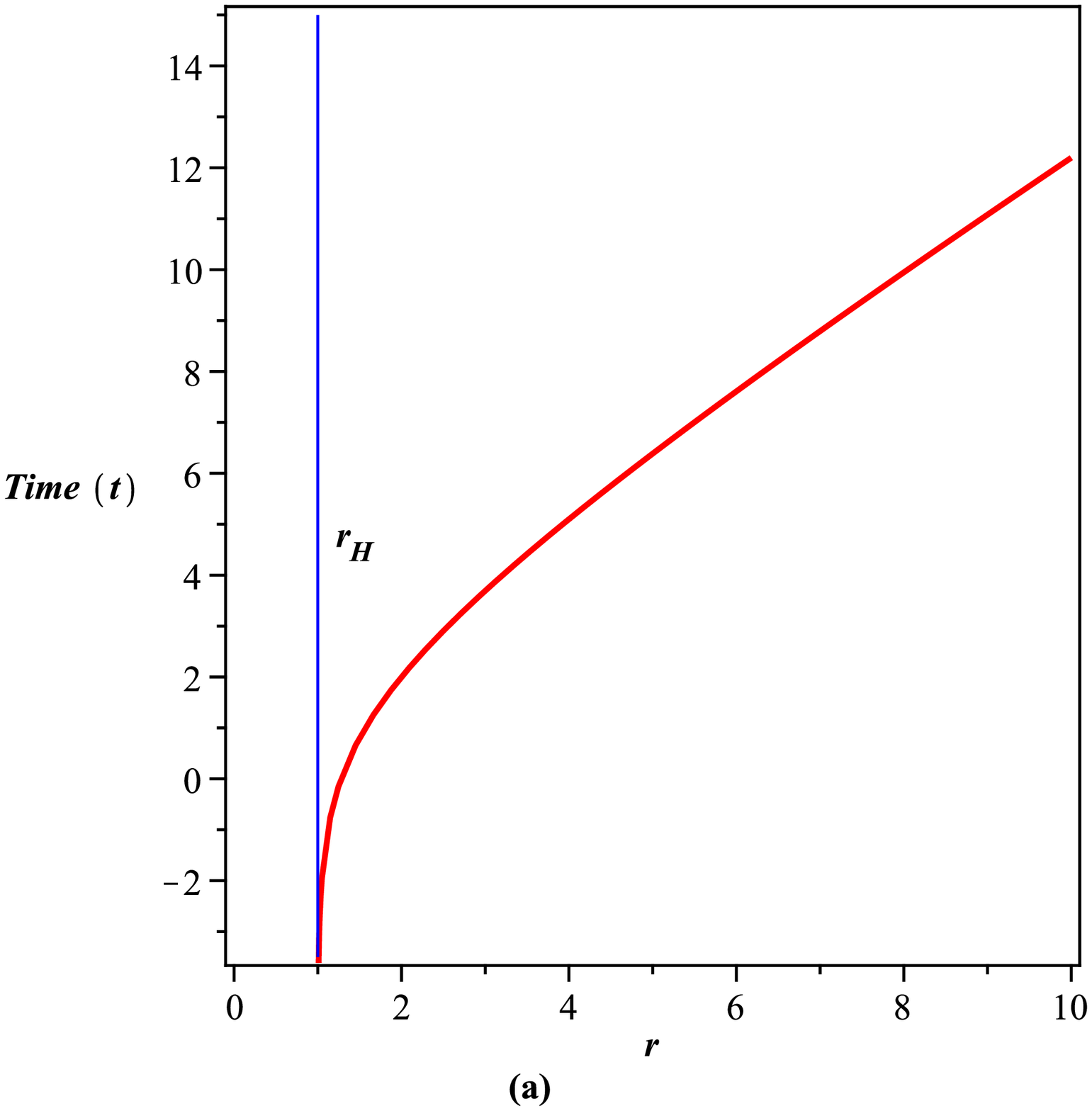}
\includegraphics[width=6.5cm,height=6cm]{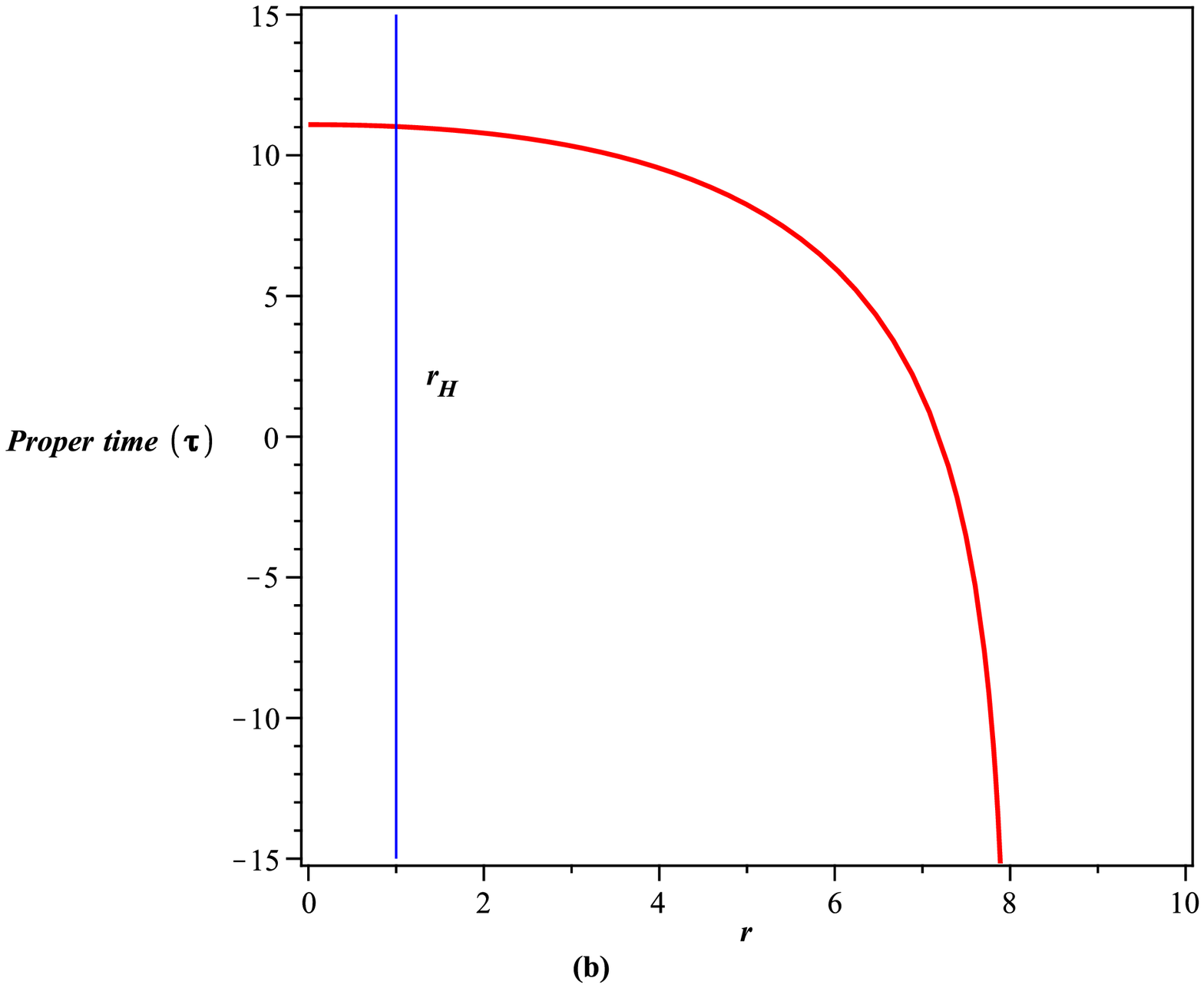}}
\caption{Radial motion of an incoming photon, with $E$ = 1 and $Q$ = 2; for (a) coordinate time; (b) proper time. Vertical solid line represents the position of event horizon ($r_H$) at $r$ = 2m in both (a) and (b).}
\protect\label{t_r}
\end{figure}

\noindent The radial geodesic corresponds to the trajectory for incoming photon with zero angular momentum (i.e. $L$ = 0).
For radial null geodesics the effective potential given in eq.(\ref{eqn:eff_potn_M}) leads to,
\begin{eqnarray}
V_{eff} = - \frac{E^{2}}{m^2 r^2}(m r - Q^2)^2 ,
\label{eqn:eff_potn_rad_M}
\end{eqnarray}
and the radial velocity given in eq.(\ref{eqn:vel_M}) reduces to,
\begin{eqnarray}
\dot{r} = \pm \frac{E (m r - Q^2)}{m r}.
\label{eqn:rt_tau_radial_M}
\end{eqnarray}
In order to obtain the coordinate time $t$ as a function of radial distance $r$, we divide eq.(\ref{eqn:first_int_t}) by eq.(\ref{eqn:rt_tau_radial_M})which leads to,
\begin{equation}
\frac{dt}{dr} = \pm \frac{r}{(r - 2 m)} .
\label{eqn:radial_t_M}
\end{equation}
On integrating eq.(\ref{eqn:radial_t_M}),
\begin{equation}
t = \pm [r + 2 m \ln(r - 2 m)] + constant.
\label{eqn:radial_t_sol_M}
\end{equation}
In the limit r $\longrightarrow 2 m$, $t \longrightarrow - \infty$.
Similarly the proper time $\tau$ can be obtained as a function of radial distance $r$ and on integrating the eq.(\ref{eqn:rt_tau_radial_M}),
\begin{equation}
\tau = \pm \frac{1}{E}\left(r+\frac{Q^2 \ln(-m r + Q^2)}{m}\right)+ constant ,
\label{eqn:radial_tau_sol_M}
\end{equation}
when r $\longrightarrow 2 m$, $\tau \longrightarrow \pm \frac{1}{E}\left(2 m + \frac{Q^2 \ln(- 2 m^2 + Q^2)}{m}\right) $.
The radial geodesics pass through the horizon in a finite proper time, while it takes an infinite time to reach the horizon as shown in fig.(\ref{t_r}a) and fig.(\ref{t_r}b). We will use $m$ = 0.5, throughout afterwards for the purpose of numerical computations.
\subsection{Geodesics with angular momentum}
\noindent Now, let us discuss the null geodesics with angular momentum (i.e. L $\neq$ 0).
The effective potential for non-radial null geodesics is given by,
\begin{equation}
V(r) = L^2 \left[\frac{(r - 2 m) (m r - Q^2)}{m r^4}-\frac{(m r - Q^2)^2}{D^2 m^2 r^2}\right] .
\label{eqn:potential-non-radial}
\end{equation}
In order to study the nature of the effective potential for null geodesics, we have considered a specific set of parameters as depicted in Table \ref{table:t1} alongwith the different conditions for the impact parameter. $D_c$ is the value of impact parameter corresponding to the unstable circular photon orbit.\\

\begin{table}[ht]
\caption{Different conditions on charge and mass parameters. }
\centering
\begin{tabular}{c c c c }
 & Case & Mass (m) & Charge (Q) \\ [0.5ex]
\hline \\
(a) & $Q^2\approx 2m^2$ & 0.5 & 0.7 \\[1ex]
(b) & $Q^2>2m^2$ & 0.5 & 1 \\[1ex]
(b) & $Q^2<2m^2$ & 0.5 & 0.4 \\[1ex]
\hline
\end{tabular}
\label{table:t1}
\end{table}
\begin{figure}[h!]
\centerline{\includegraphics[width=6cm,height=6.5cm]{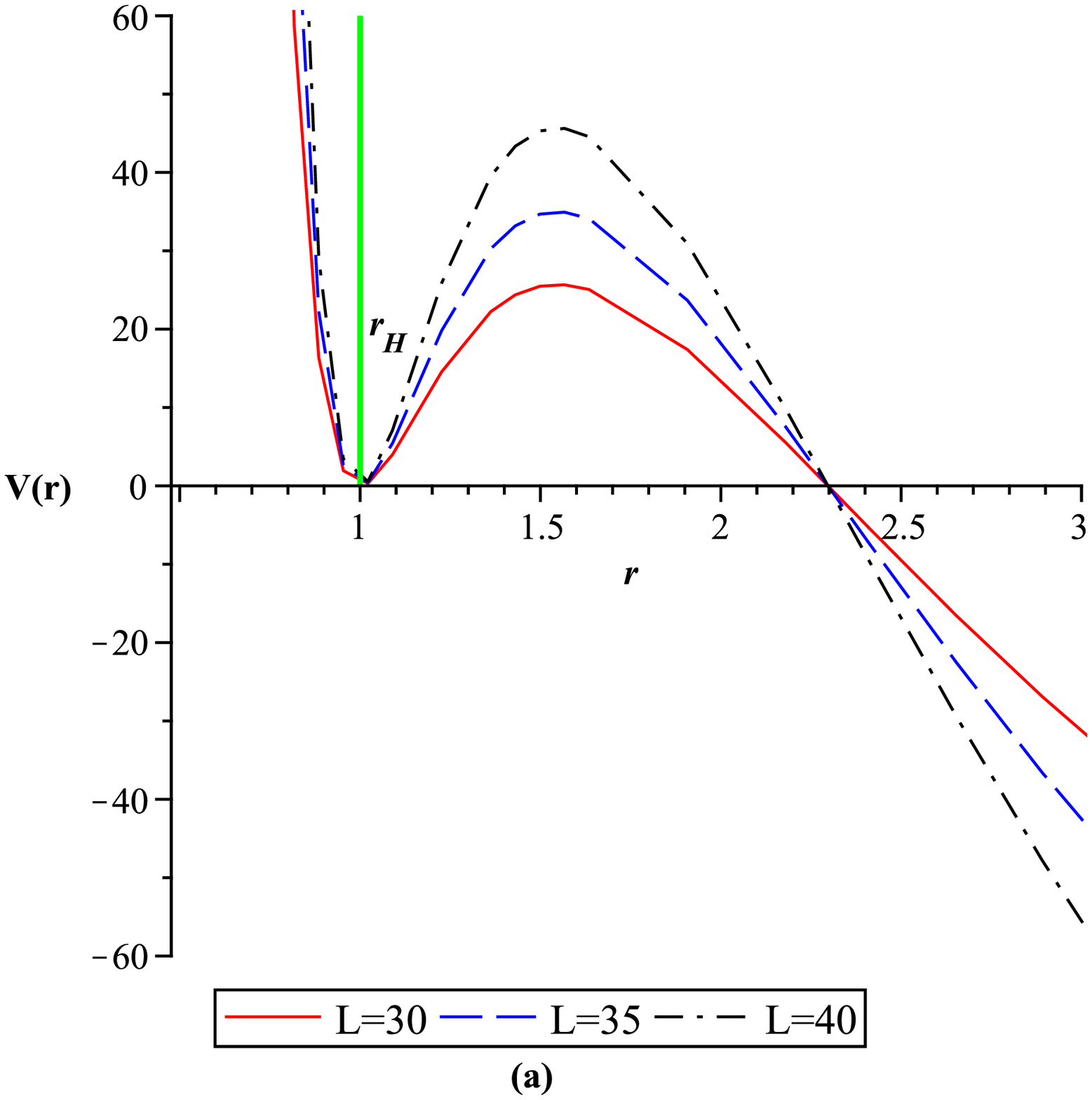}
\includegraphics[width=6cm,height=6.5cm]{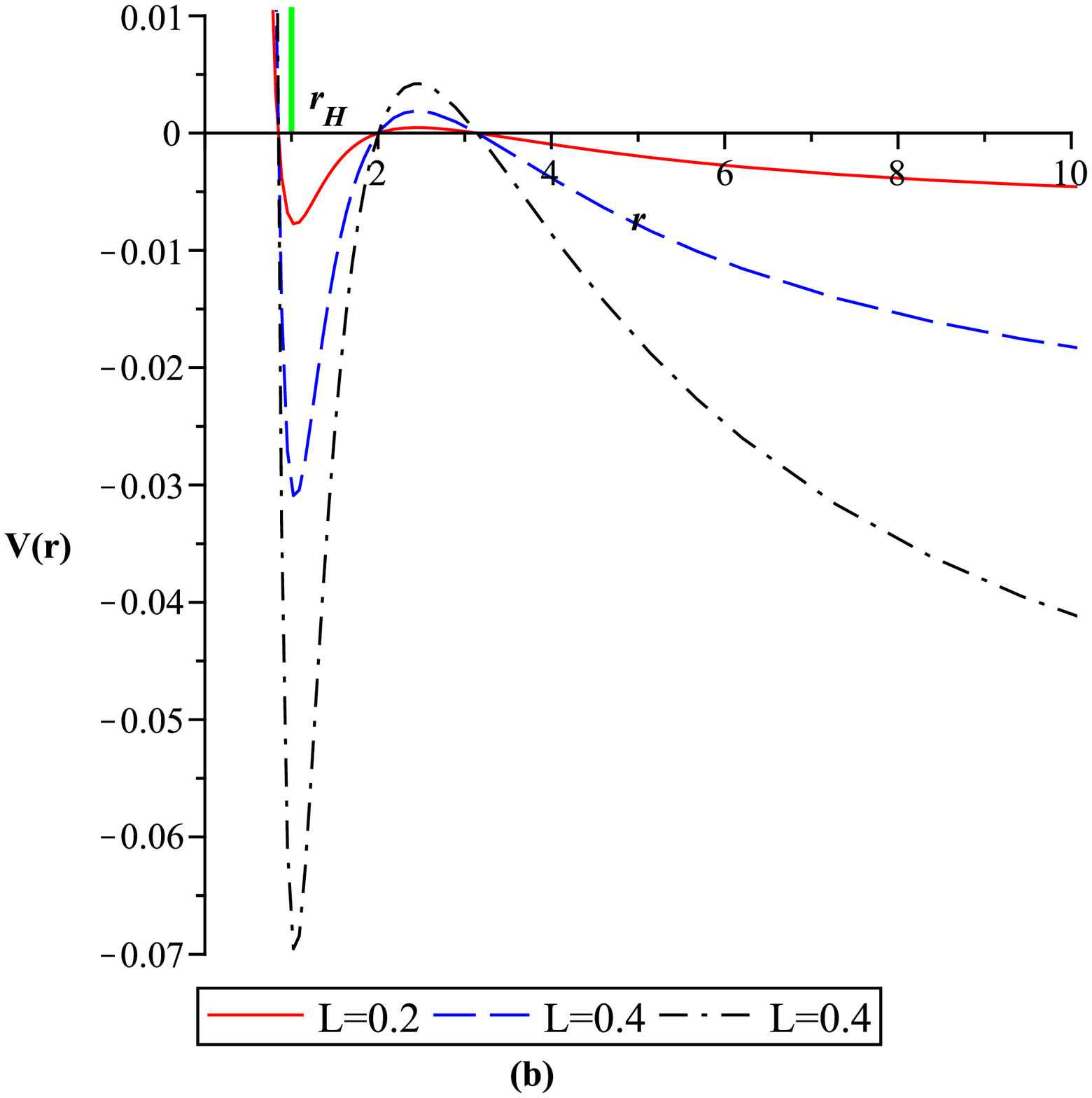}
\includegraphics[width=6cm,height=6.5cm]{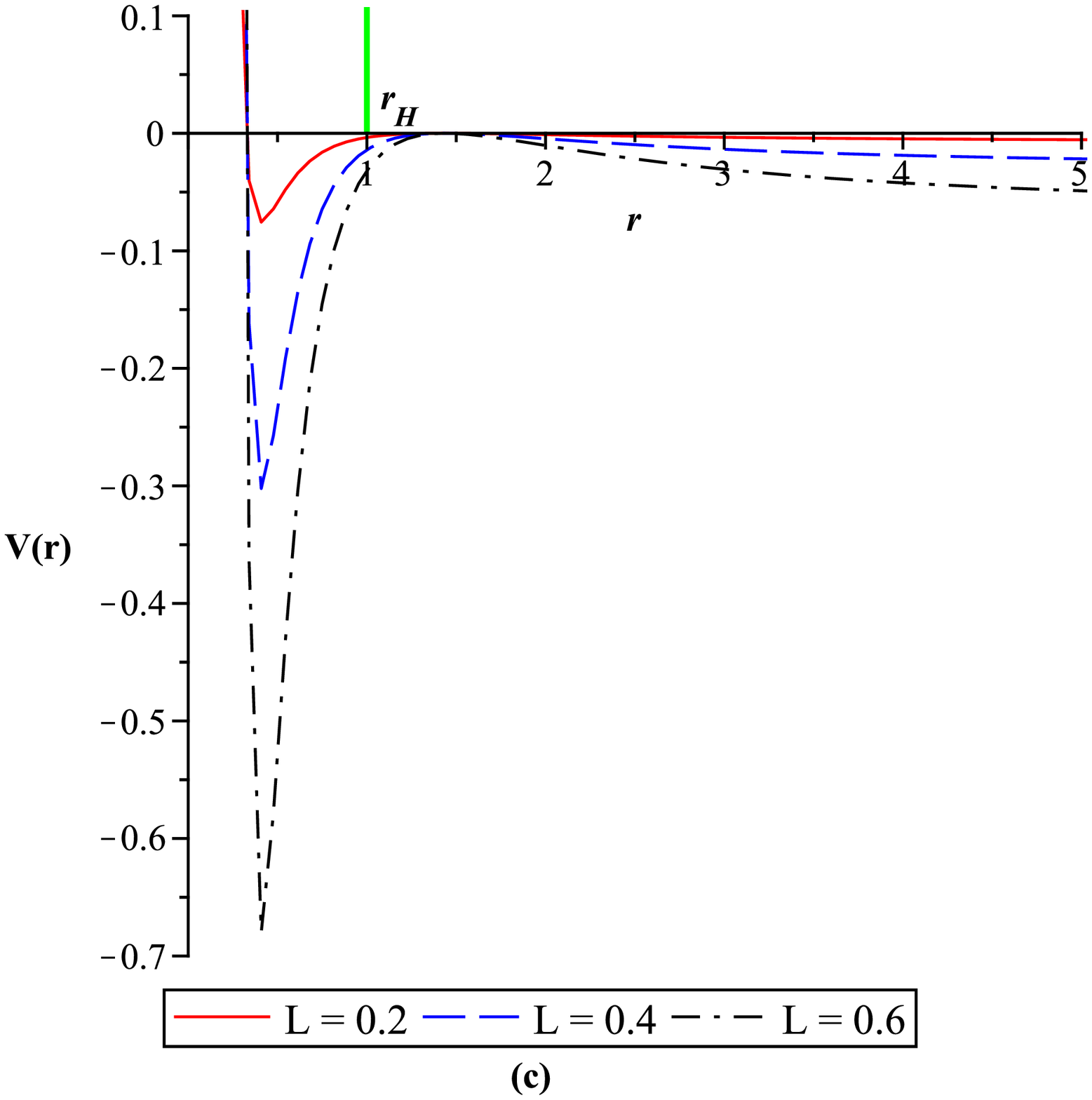}}
\caption{$D = D_c$ : Variation of effective potential with angular momentum for three different cases as mentioned in Table I, where $r_H$ represents the position of the event horizon.}
\protect\label{f1}
\end{figure}
\begin{figure}[h!]
\centerline{\includegraphics[width=6cm,height=6.5cm]{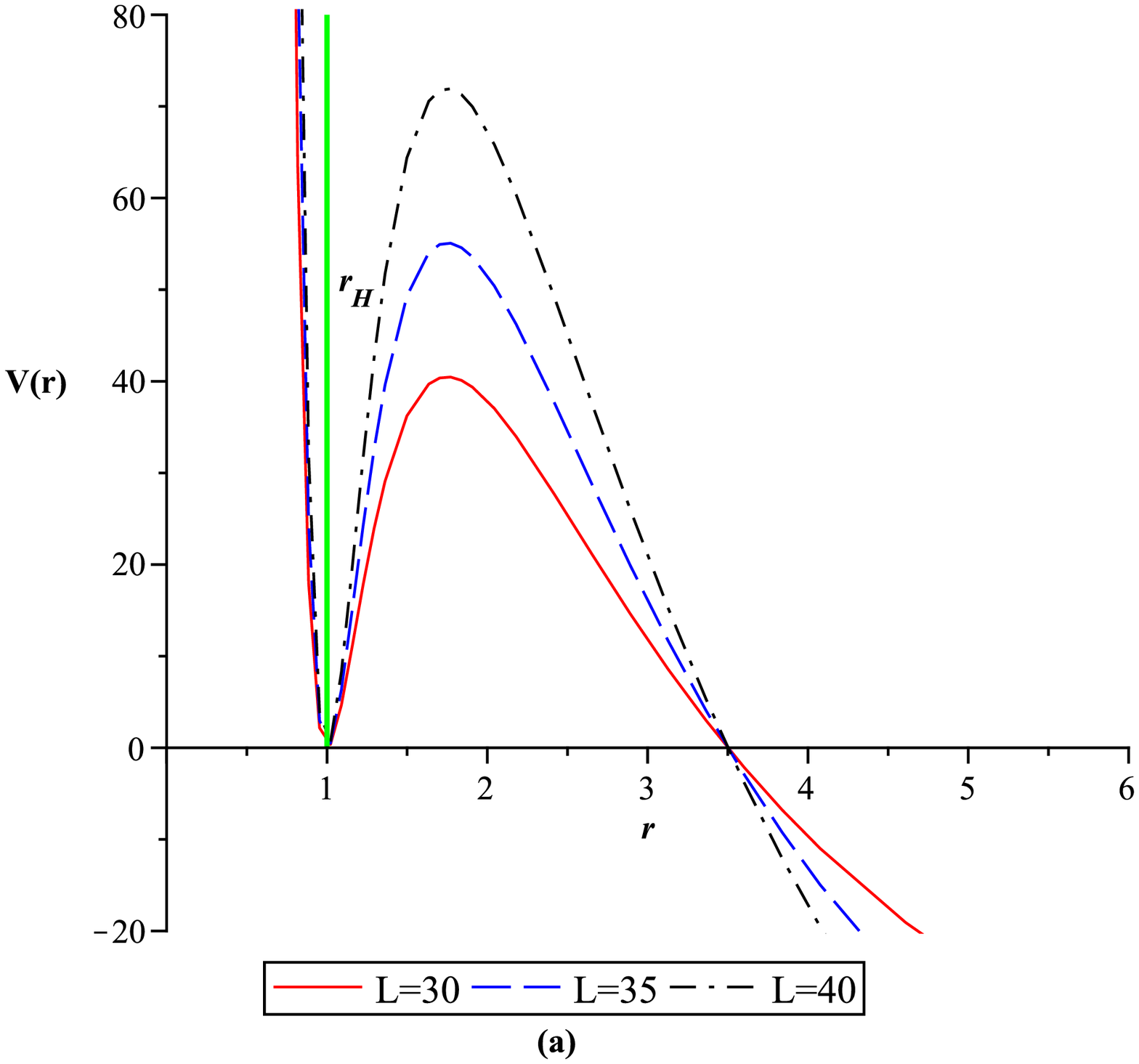}
\includegraphics[width=6cm,height=6.5cm]{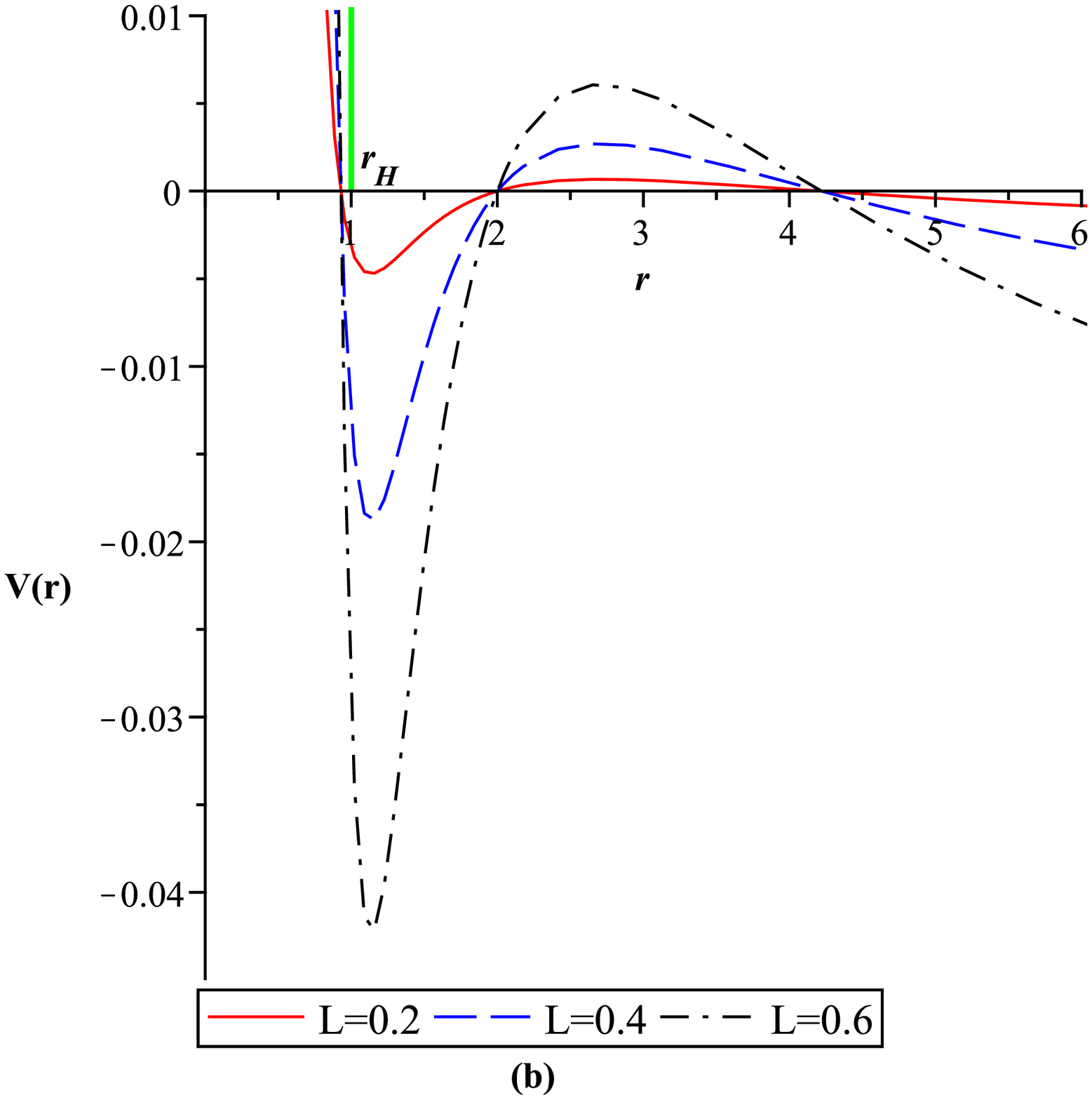}
\includegraphics[width=6cm,height=6.5cm]{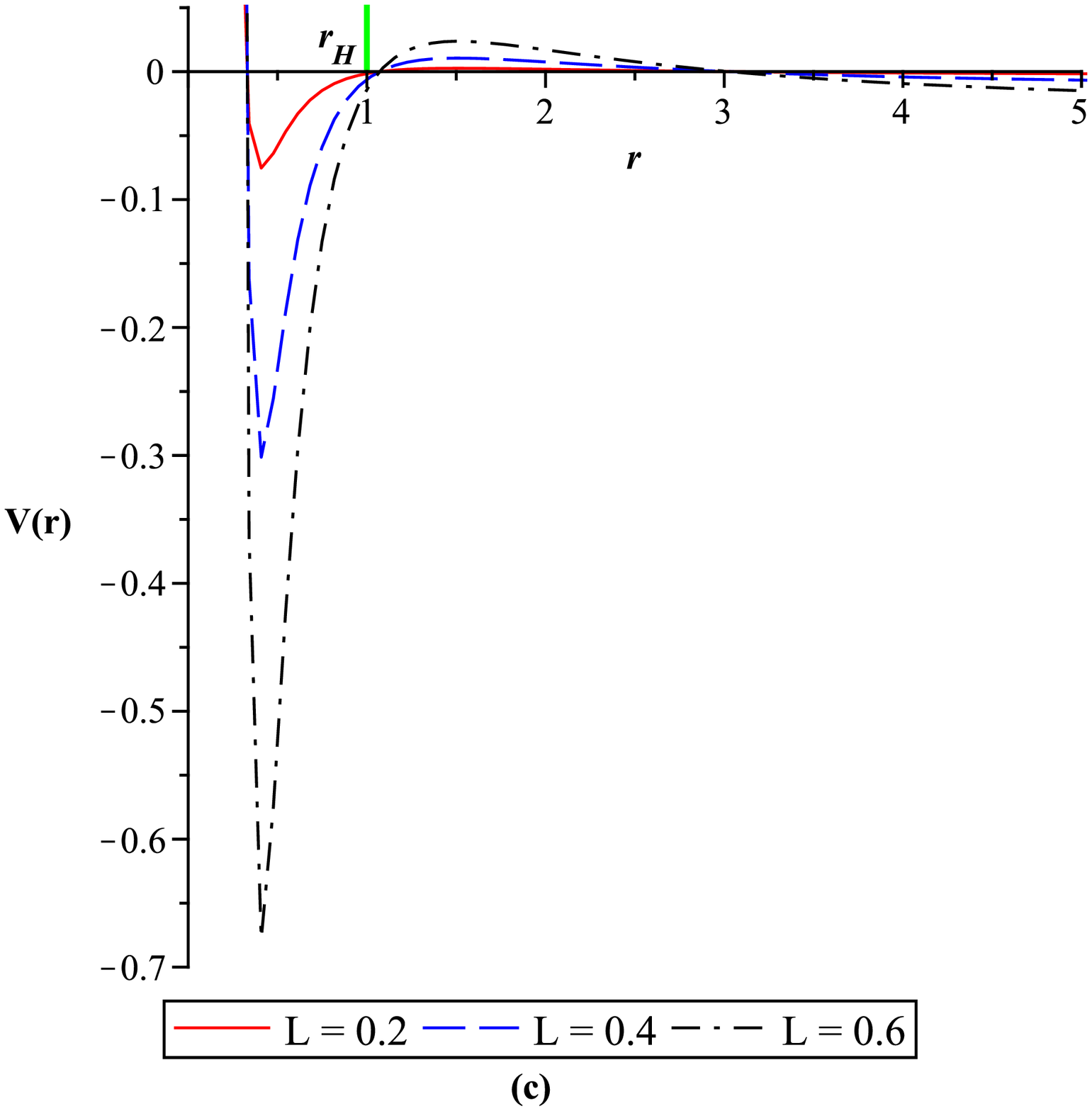}}
\caption{$D > D_c$: Variation of effective potential with angular momentum for three different cases mentioned in Table I. $r_H$ is the position of the event horizon.}
\protect\label{f2}
\end{figure}
\begin{figure}[h!]
\centerline{\includegraphics[width=6cm,height=5.5cm]{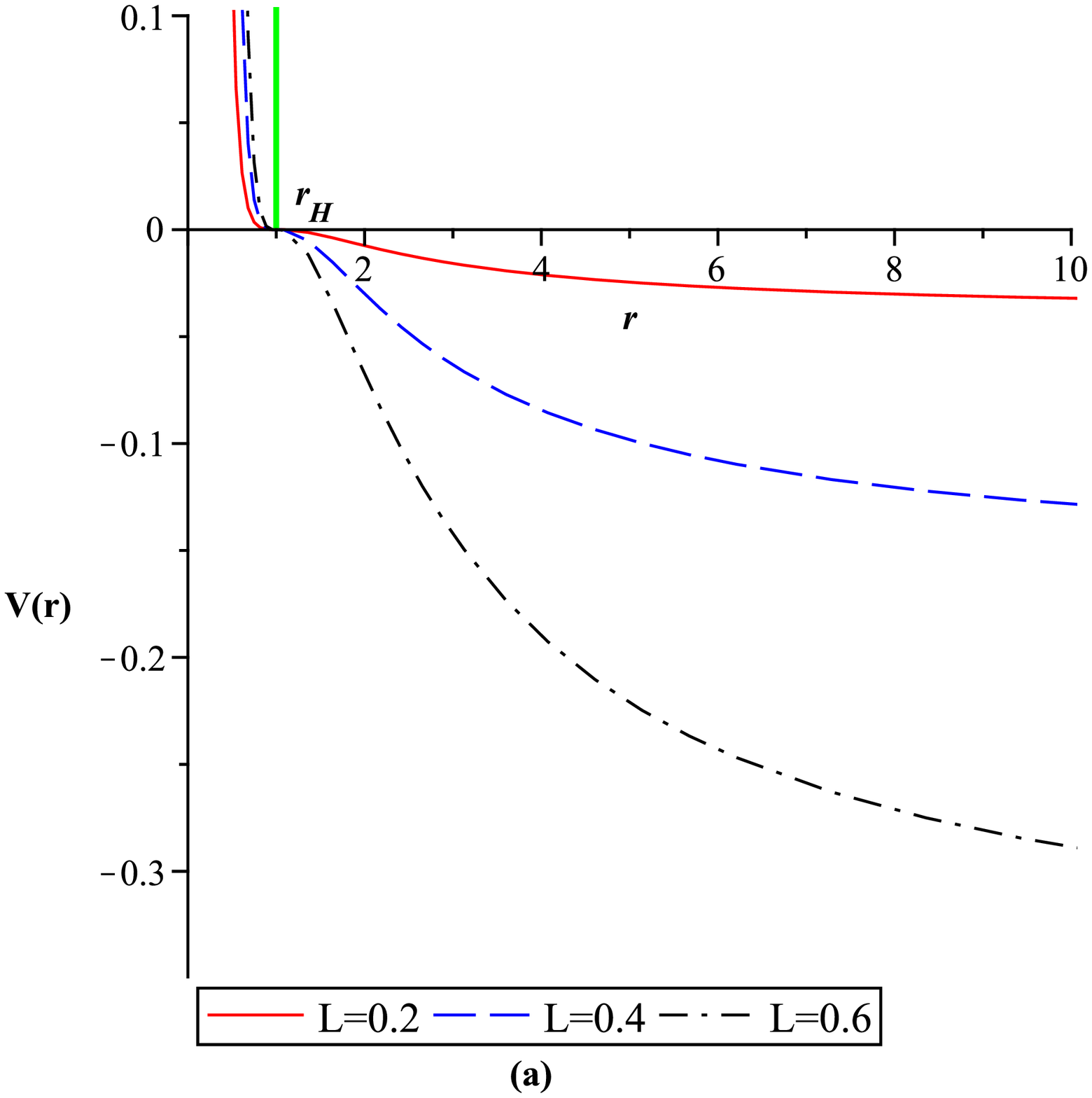}
\includegraphics[width=6cm,height=5.5cm]{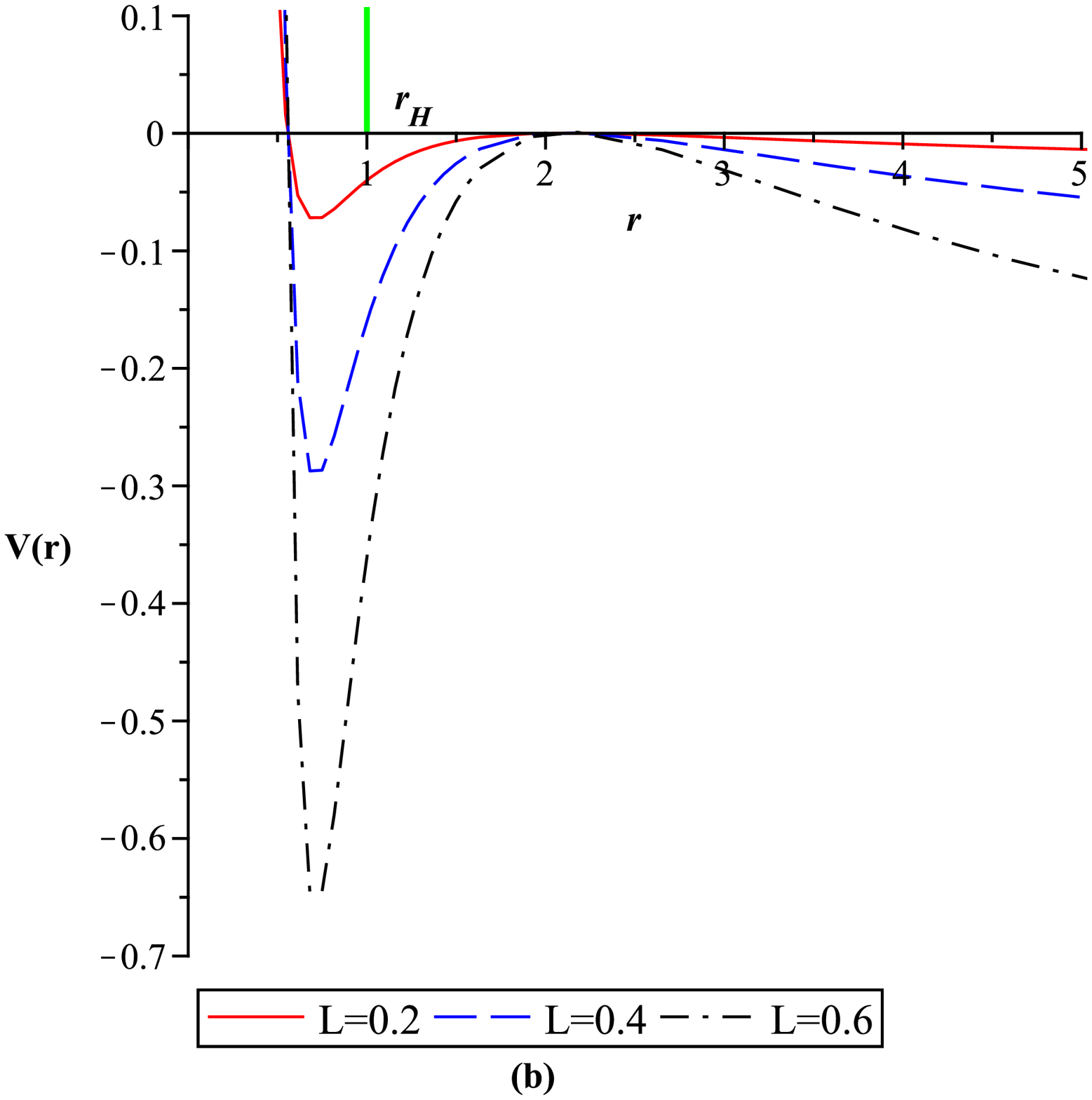}
\includegraphics[width=6cm,height=5.5cm]{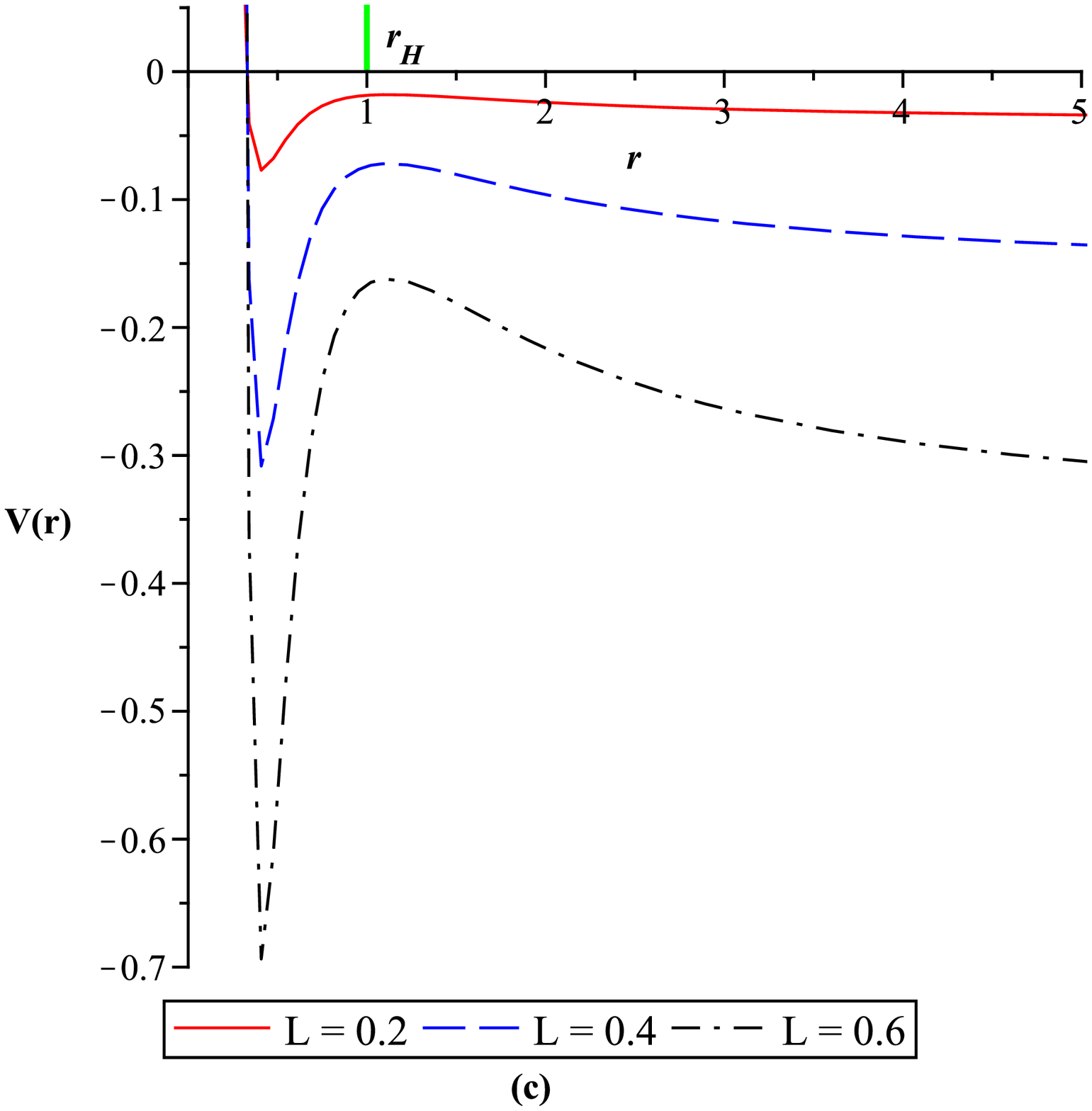}}
\caption{$D < D_c$: Variation of effective potential with angular momentum for three different cases mentioned in Table I, where $r_H$ denotes the event horizon.}
\protect\label{f3}
\end{figure}
\newpage
\noindent The variation of effective potential is presented in fig.(\ref{f1}), fig.(\ref{f2}) and fig.(\ref{f3}) with a finite value for angular momentum $L$ of the incoming photon while other parameters are fixed.
The nature of effective potential is qualitatively similar for the cases $D=D_c$ and $D>D_c$ as the height of effective potential increases with increasing the value of angular momentum see figs.(\ref{f1}) and (\ref{f2}).
While this nature is found to be reversed for the case $D<D_c$, as seen in fig.(\ref{f3}).
The difference in the corresponding orbit structures due to this difference in the nature of effective potential is discussed in detail later in this section.\\
\begin{figure}[h!]
\centerline{\includegraphics[width=6cm,height=4.5cm]{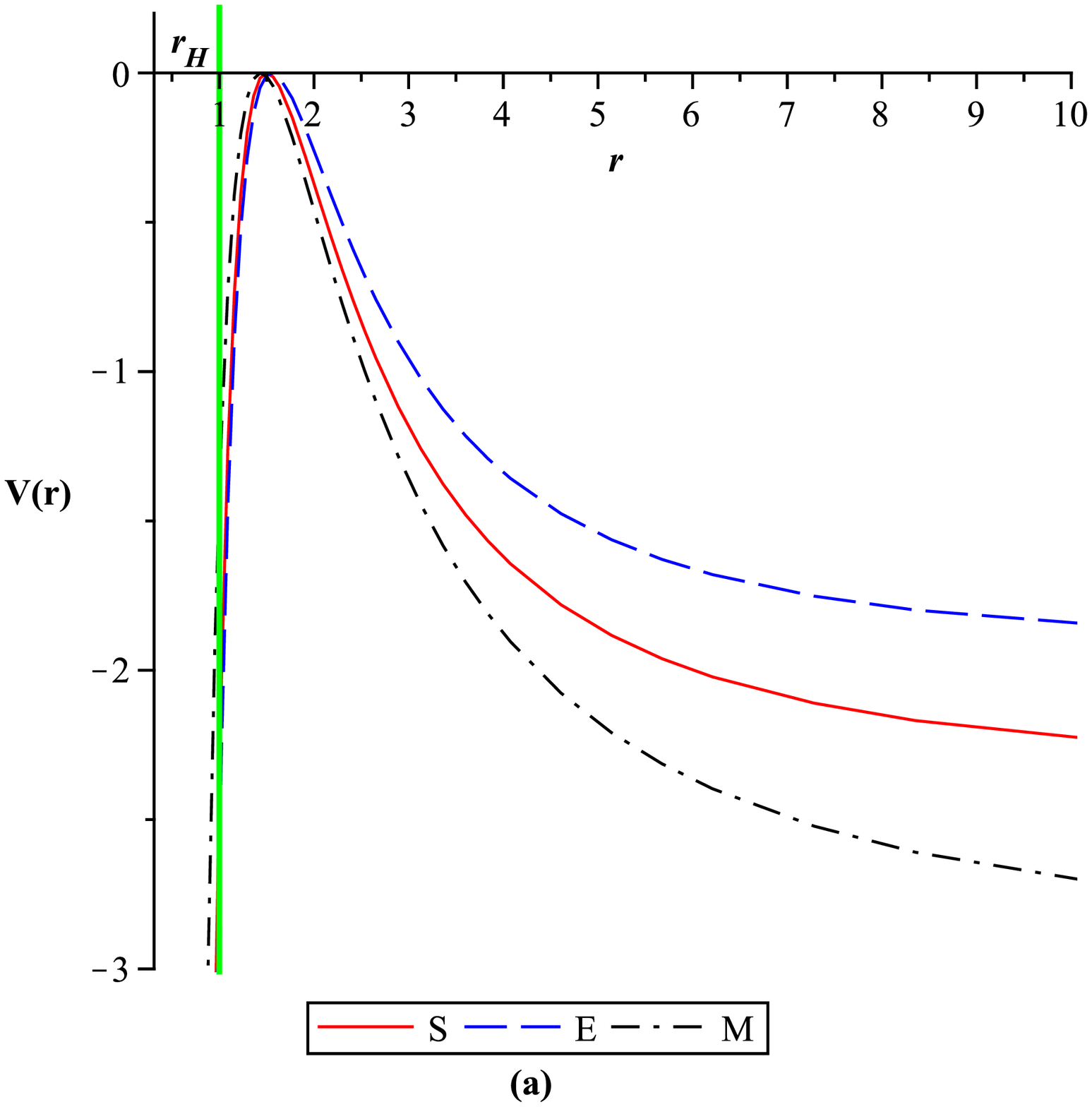}
\includegraphics[width=6cm,height=4.5cm]{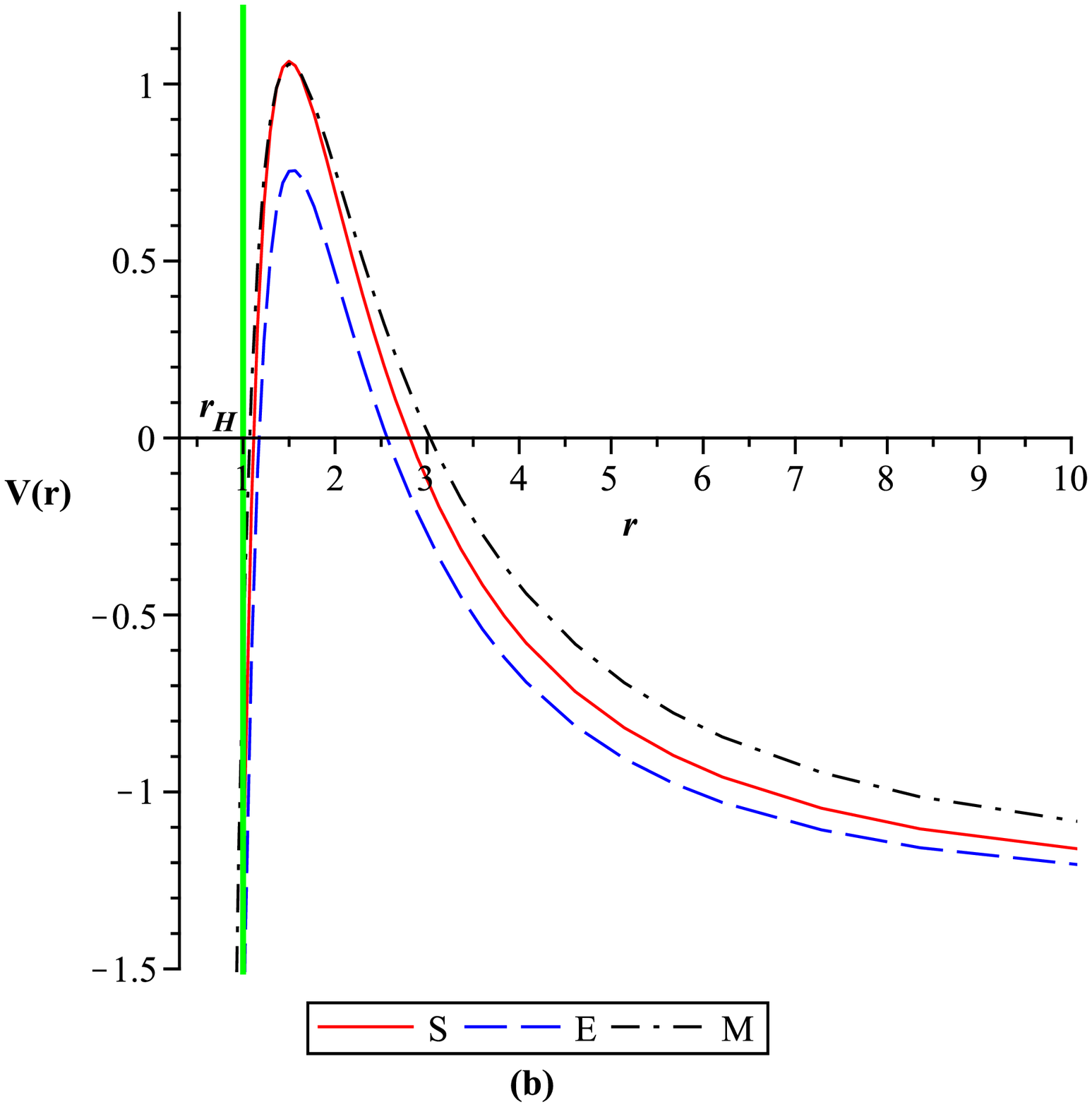}
\includegraphics[width=6cm,height=4.5cm]{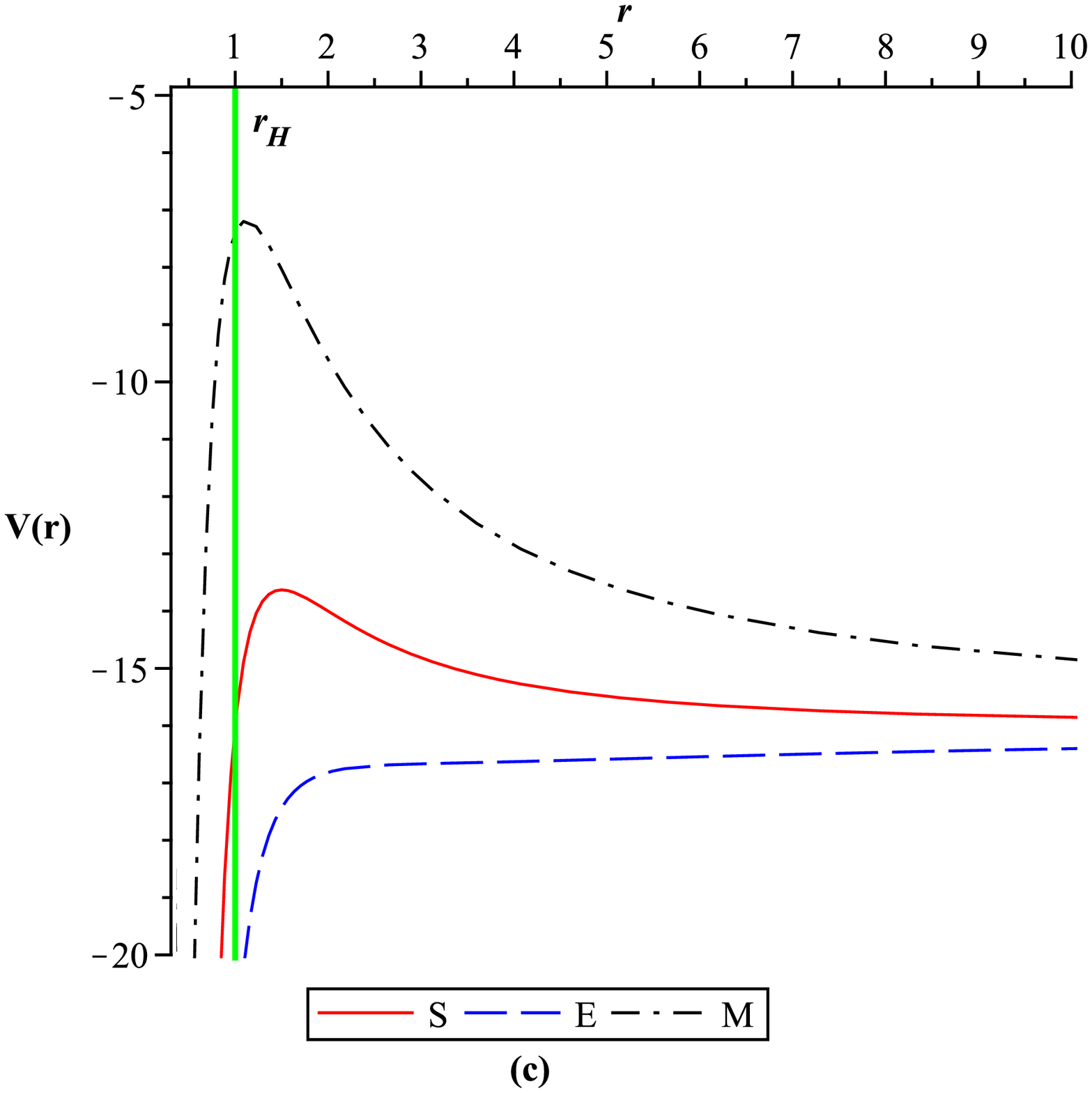}}
\caption{Comparative view of the nature of effective potential for three different cases of impact parameter, (a)$D$ = $D_c$, value of $D_c$ for $S$, $E$ and $M$ is 2.5980, 2.8852 and 2.2974 respectively; \\(b) $D$ ( = 3.5) $> D_c$; (c) $D$ ( = 1) $< D_c$, with $Q$ = $\alpha$ = 0.4 and $L$ = 4. $r_H$ (represented by vertical line) is the position of event horizon; $S$, $E$ and $M$ represent the SBH, ECBH and MCBH respectively.}
\protect\label{com_v}
\end{figure}\\
\noindent From the representative plots of the effective potential presented in fig.(\ref{com_v}), one can observe the shift of the potential energy curve for MCBH and ECBH from the corresponding curve for SBH, with different conditions of impact parameter.
\subsection{Analysis of photon orbits}
\begin{figure}[h]
\centerline{\includegraphics[width=6cm,height=6cm]{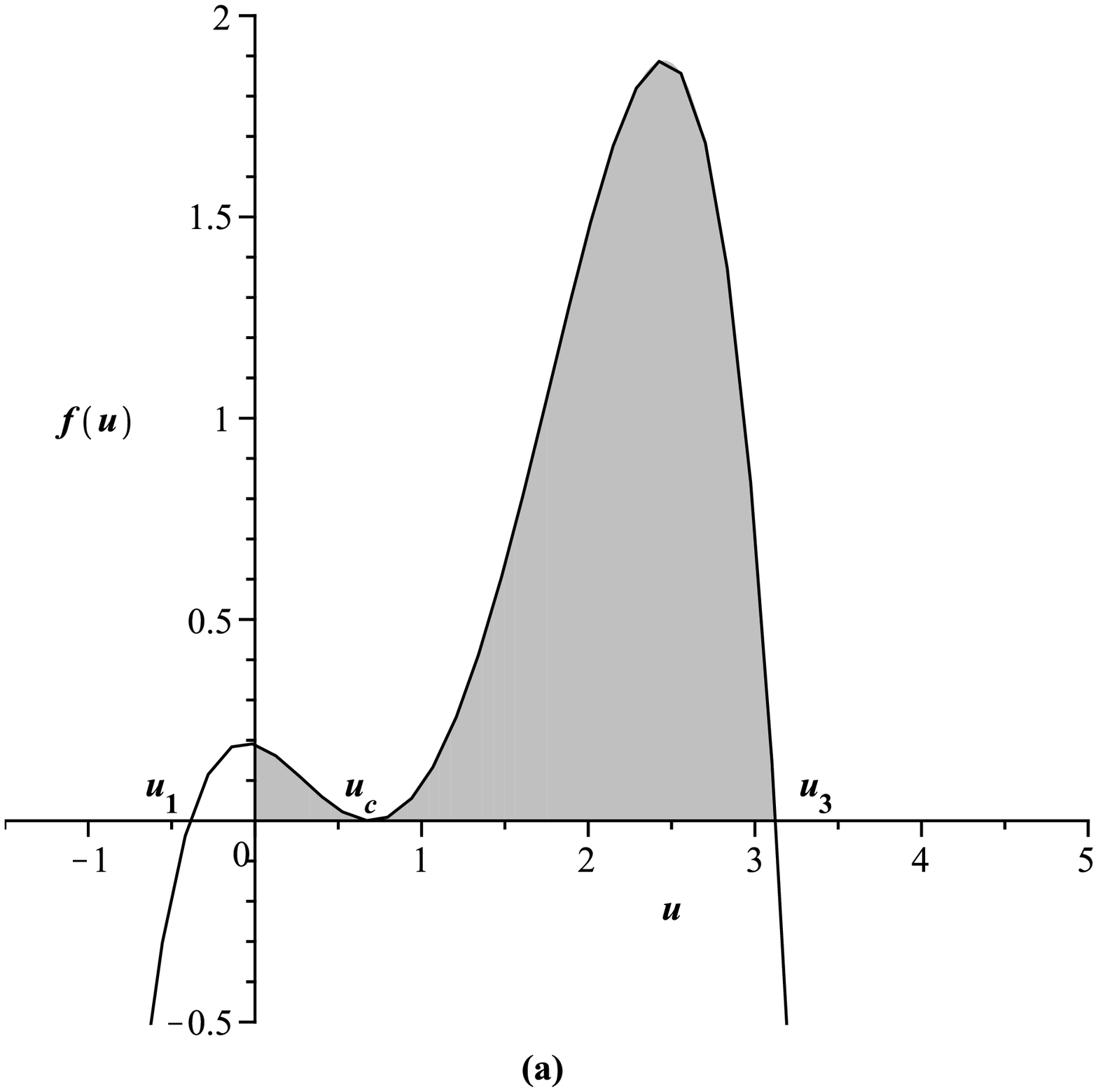}
\includegraphics[width=6cm,height=6cm]{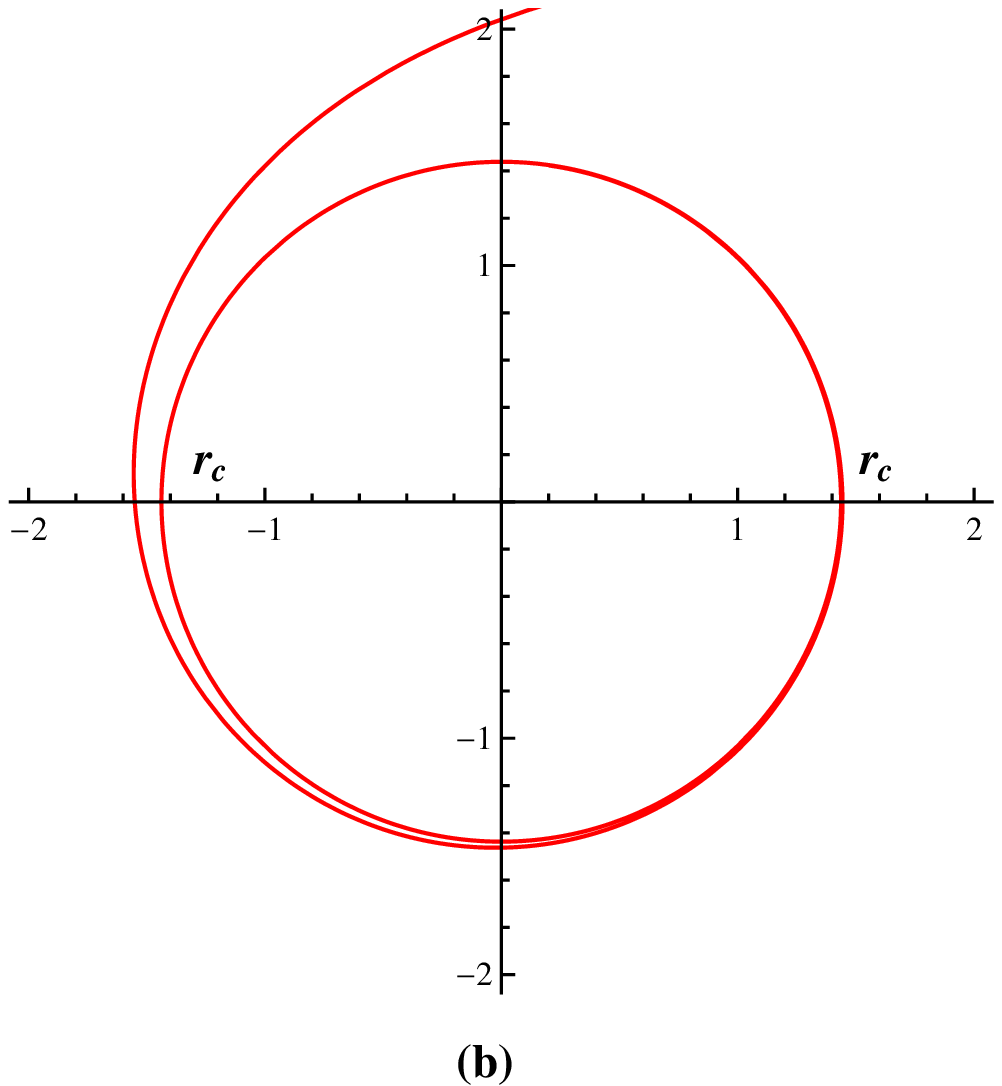}
\includegraphics[width=6cm,height=6cm]{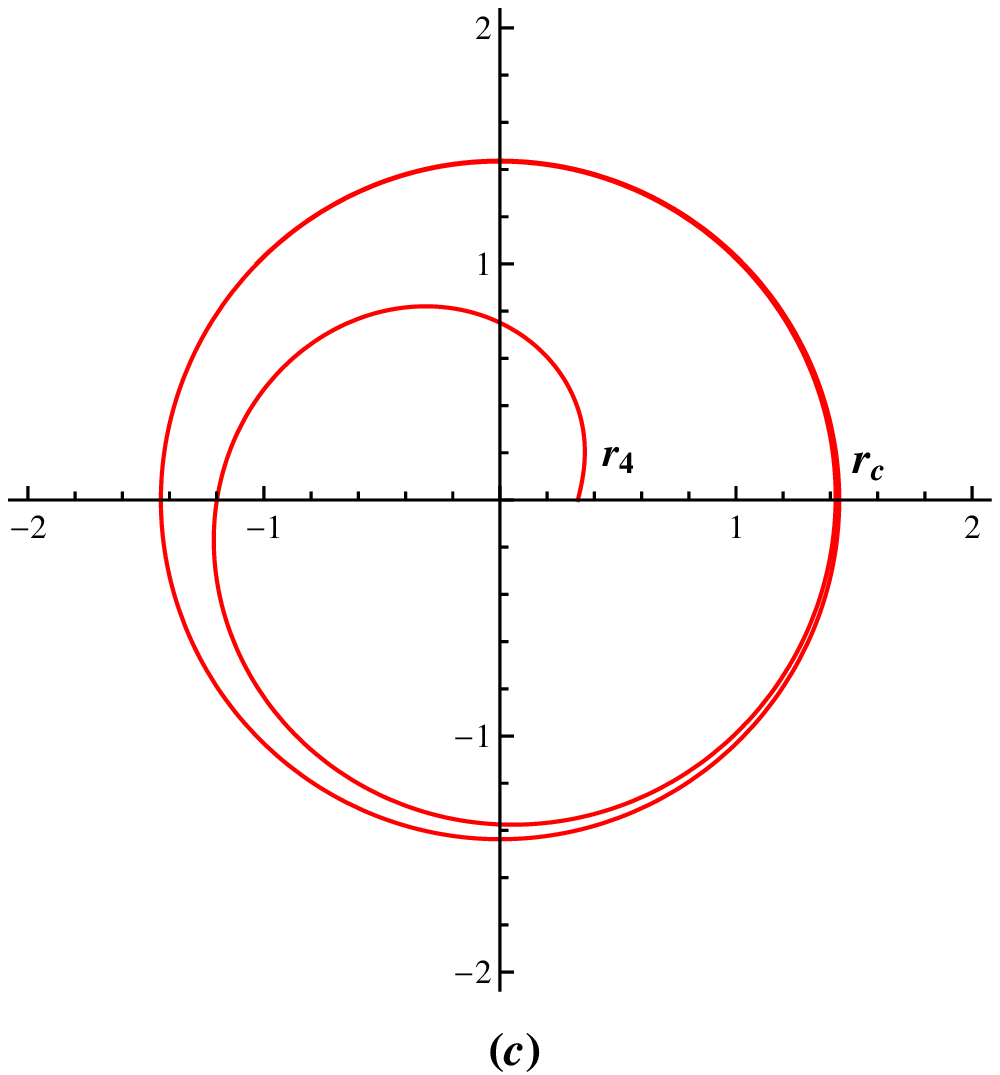}}
\caption{(a) $f$(u) represents the shaded allowed region for the motion of photons with a positive root at points $\bf{u_4}$ and a degenerate root at $\bf{u_c}$ (where the {\it unstable circular orbit} exists), with D = 2.29 and Q = 0.4; (b) Solid line represents the unstable circular orbit with radius $r_c = 1.43$ for a photon starting from infinity; (c) Solid line represents a {\it bound orbit} spiralling around the circular radius $r_c$ starting form the singularity. In fig.(a), $u$ is used as a coordinate while $r$ is used as a coordinate in figs. (b) and (c).}
\label{f4}
\end{figure}

\noindent To analyse the angular motion of photons, the corresponding orbit equation can be obtained by eliminating parameter $\tau$ from eqn.(\ref{eqn:vel_M}) and eqn.(\ref{eqn:first_intg_M}) as,
\begin{equation}
\frac{d \phi}{d r} = \frac{L}{r^2} \frac{1}{\sqrt{ - V_{eff}}} ,
\end{equation}
which can be recast as,
\begin{equation}
\left(\frac{d r}{d \phi}\right)^2 = f(r) ,
\label{eqn:fr_M}
\end{equation}
where
\begin{equation}
f(r) = r^4 \left[- 2 \frac{Q^2}{r^4} + J \frac{1}{r^3} + K \frac{1}{r^2}
- 2 \frac{Q^2}{D^2 m}\frac{1}{r} + \frac{1}{D^2}\right],
\nonumber
\label{f-mag}
\end{equation}
 with
\begin{equation}
J = \frac{Q^2}{m} + 2 m\hspace{3mm}and \hspace{3mm} K = \frac{Q^4}{D^2 m^2} - 1.
\nonumber
\end{equation}
In order to investigate the nature of orbits, we use a variable as $u = \frac{1}{r}$, so the eq.(\ref{eqn:fr_M}) leads to,
\begin{equation}
\left(\frac{d u}{d \phi}\right)^2 = f(u),
\label{eqn:u-phi_M}
\end{equation}
where $f(u)$ is given as,
\begin{equation}
f(u) = - 2 Q^2 u^4 + J u^3 + K u^2 - 2 \frac{Q^2 u}{D^2 m} + \frac{1}{D^2} .
\label{f-u-mag}
\end{equation}
The null geodesics depends on the roots of the equation $f(u) = 0$ \cite{cha83, fer12}.
All the possible orbits for null geodesics in view of the different values of impact parameter $D$ can now be discussed accordingly.\\
\vspace{3mm}\\
\noindent\textbf{Case (a) : For $D$ = $D_c$}\\
Using the condition for circular orbit \cite{fer12},
\begin{equation}
\dot{r} = 0 \Rightarrow V_{eff} (r) = 0 ,
\label{eqn:cir_con1}
\end{equation}
along with $V^{\prime}_{eff}= 0$\footnote{where $\prime$ denotes the differentiation w.r.t. $r$}, one can obtain the radius of the circular orbit as,
\begin{equation}
r_{\pm} = \frac{(Q^2 + 6 m^2) \pm \sqrt{Q^4 - 20 m^2 Q^2 + 36 m^4}}{4 m} .
\end{equation}
A stable (unstable) circular orbit requires $ V^{\prime\prime}(r) > 0 (<0)$, which corresponds to minimum (maximum) of the effective potential. Therefore the radius of unstable circular orbit ($r_c$) occurs at,
\begin{equation}
r_{c} =  \frac{Q^2 + 6 m^2 + \sqrt{Q^4 - 20 m^2 Q^2 + 36 m^4}}{4 m}.
\label{eqn:rc}
\end{equation}
Since with $Q \rightarrow 0$, the MCBH reduce to the SBH and the corresponding value of $r_c$ given in eq.(\ref{eqn:rc}) also reduces to $3 m$ which is the exact value of the radius of unstable circular orbit for SBH. Here, $r_c$ is independent of the parameters $E$ and $L$.
For circular orbits, the impact parameter is given as,
\begin{equation}
 D_c^2 = \frac{L^2_{c}}{E^2_{c}}=\frac{r_c^2 (m r_c -Q^2)}{m (r_c-2m)} .
 \label{eqn:IP_mag}
\end{equation}
Again as $Q \rightarrow 0$, $D_c^2=27 m^2$, which is the value of impact parameter for Schwarzschild BH \cite{cha83}.
Similarly the impact parameter for circular orbits for ECBH given in eqn.(\ref{eq:electric}) is given as,
\begin{equation}
 D_{c}^E = \sqrt{\frac{r_c (r_c + 2 m sinh^2\alpha)^2}{r_c - 2 m}} .
 \label{eqn:IP_electric}
\end{equation}
Now $f(u)$ given in eq.(\ref{f-u-mag}) has a real root at $u_4$ = $m/Q^2$ and hence it can also be written as,
\begin{equation}
f(u) = 2(m - Q^2 u)g(u),
\nonumber
\end{equation}
where
\begin{equation}
g(u) = u^3 - \frac{u^2}{2 m} - \frac{Q^2 u}{2 D^2 m^2} + \frac{1}{2 m D^2}.
\end{equation}
The roots of $g(u)$ are related as,
\begin{eqnarray}
u_1 + u_2 + u_3 = \frac{1}{2 m},\nonumber\\
u_1 u_2 u_3 = - \frac{1}{2 m D^2} .
\end{eqnarray}
The circular orbit exists when $f(u)$ has a real doubly degenerate root $(u_2 = u_3 = u_c)$ and hence
\begin{equation}
f(u)= 2(m - Q^2 u)(u - u_c)^2 (u - u_1)
\label{eqn:fu_c},
\end{equation}
where $u_1 = (1/2m) - 2 u_c$.\\
\newpage
\noindent
The trajectory of a photon with $D$ = $D_c$ is presented in fig.(\ref{f4}).
The orbit equation given in eq.(\ref{eqn:u-phi_M}) is integrated by using eq.(\ref{eqn:fu_c}).
Hence, the analytic solution for the circular orbit reads as,
\begin{equation}
u(\phi) = \frac{Q^4 u_c u_1^2 - 4 m A u_1 + B u_c + (- 2 m Q^2 + 2 Q^2 A)u_1 u_c}{Q^4 u_1^2 - (2 m Q^2 + 2 Q^2 A)u_1 + 4 Q^2 A u_c - 2 m A + m^2 + A^2} ,
\label{eqn:sol_orbit_M}
\end{equation}
where ,
\begin{equation}
A =\exp\left(\sqrt{2 (Q^2 u_c - m)(- u_c + u_1)}\right)(\phi + \phi_0),
\nonumber
\end{equation}
\begin{equation}
B =A^2 + 2 m A + m^2.
\end{equation}
With initial condition $u = 0$, $\phi = 0$, the constant of integration $\phi_0$ is given as,
\begin{equation}
\phi_0 = \frac{1}{\sqrt{2} (Q^2 u_c - m)(- u_c + u_1)}\ln\left(\frac{- m u_c - Q^2 u_1 u_c + 2 m u_1 - 2 (Q^2 u_c - m)(- u_c + u_1) \sqrt{ - m u_1}}{u_c}\right).
\end{equation}

\begin{figure}[h!]
\centerline{\includegraphics[width=5cm,height=7cm]{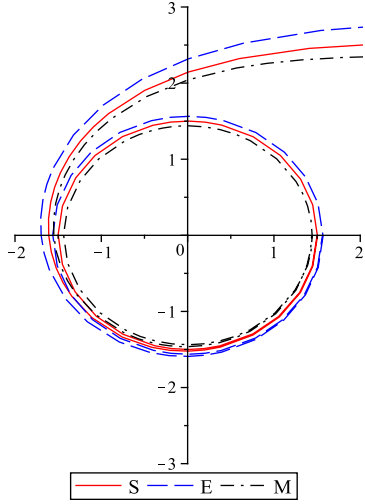}}
\caption{Comparative view of unstable circular orbits for SBH, ECBH and MCBH, represented by curves $S$, $E$ and $M$ respectively, with $Q$ = $\alpha$ = 0.4. Numerical values of the corresponding radius for unstable circular orbits are $r^S_{c}$ = 1.5, $r^E_{c}$ = 1.55 and $r^M_{c}$ = 1.43.}
\protect\label{f5}
\end{figure}
\begin{figure}[h!]
\centerline{\includegraphics[width=6cm,height=6cm]{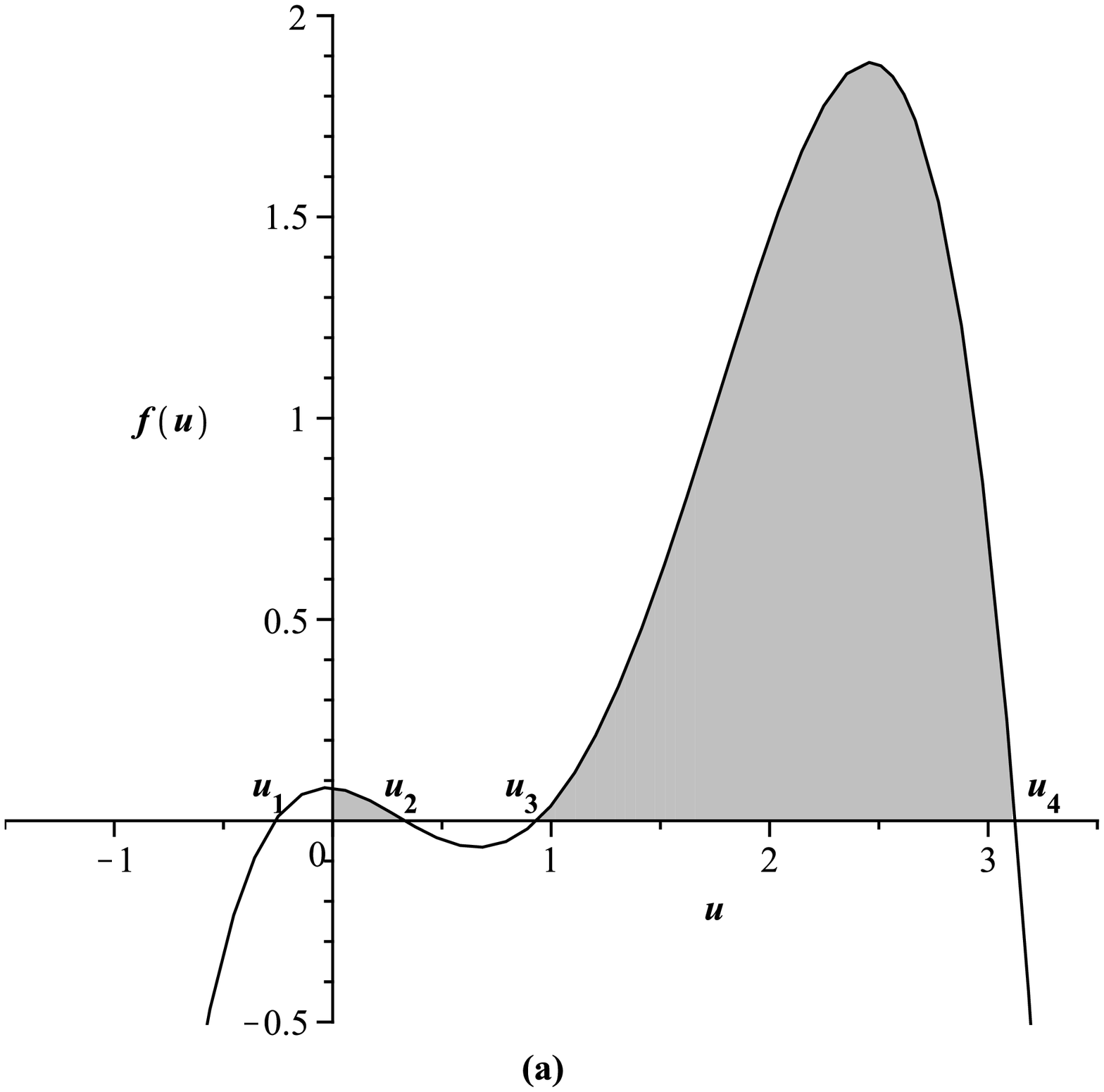}
\includegraphics[width=6cm,height=6cm]{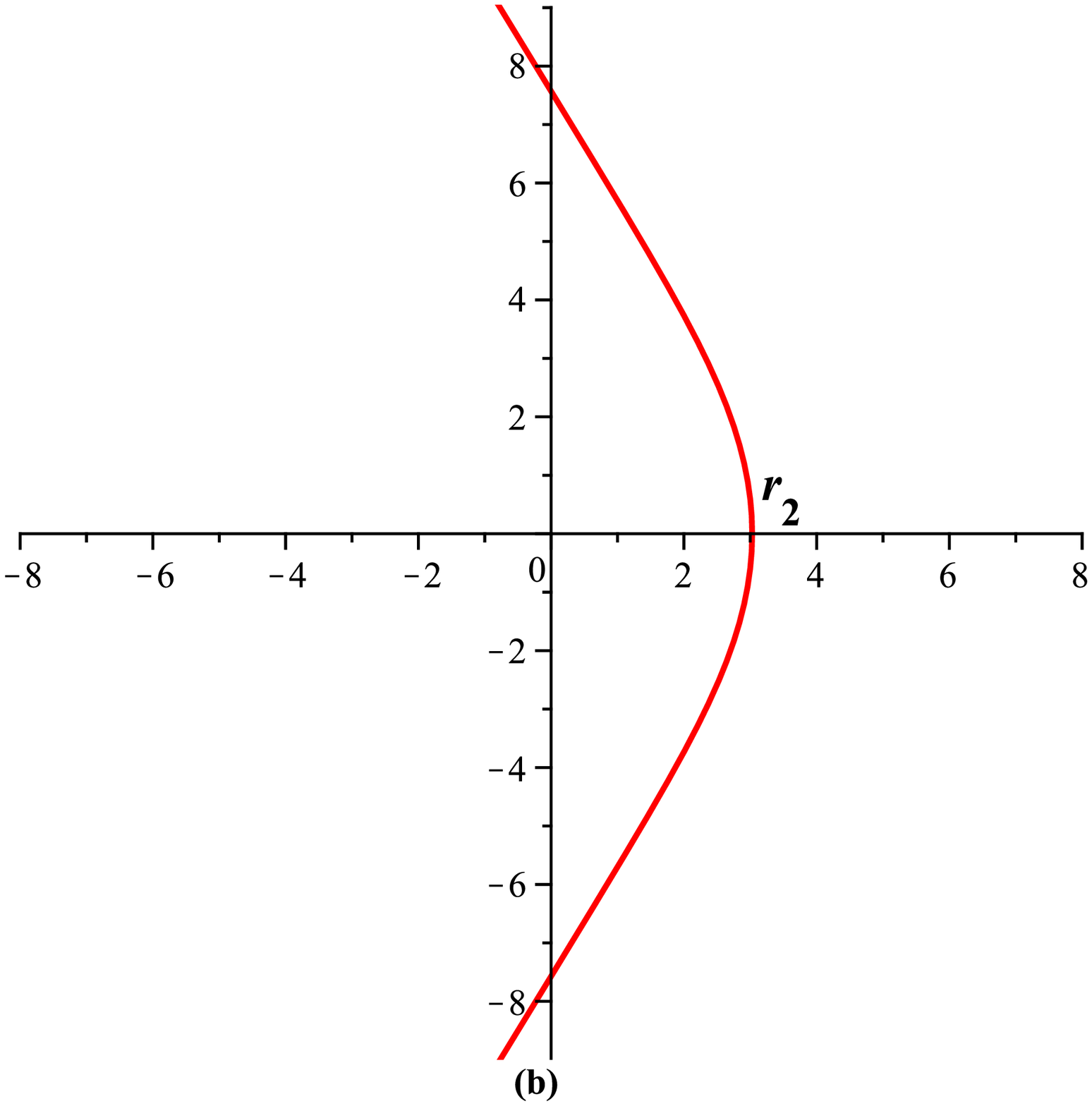}
\includegraphics[width=6cm,height=6cm]{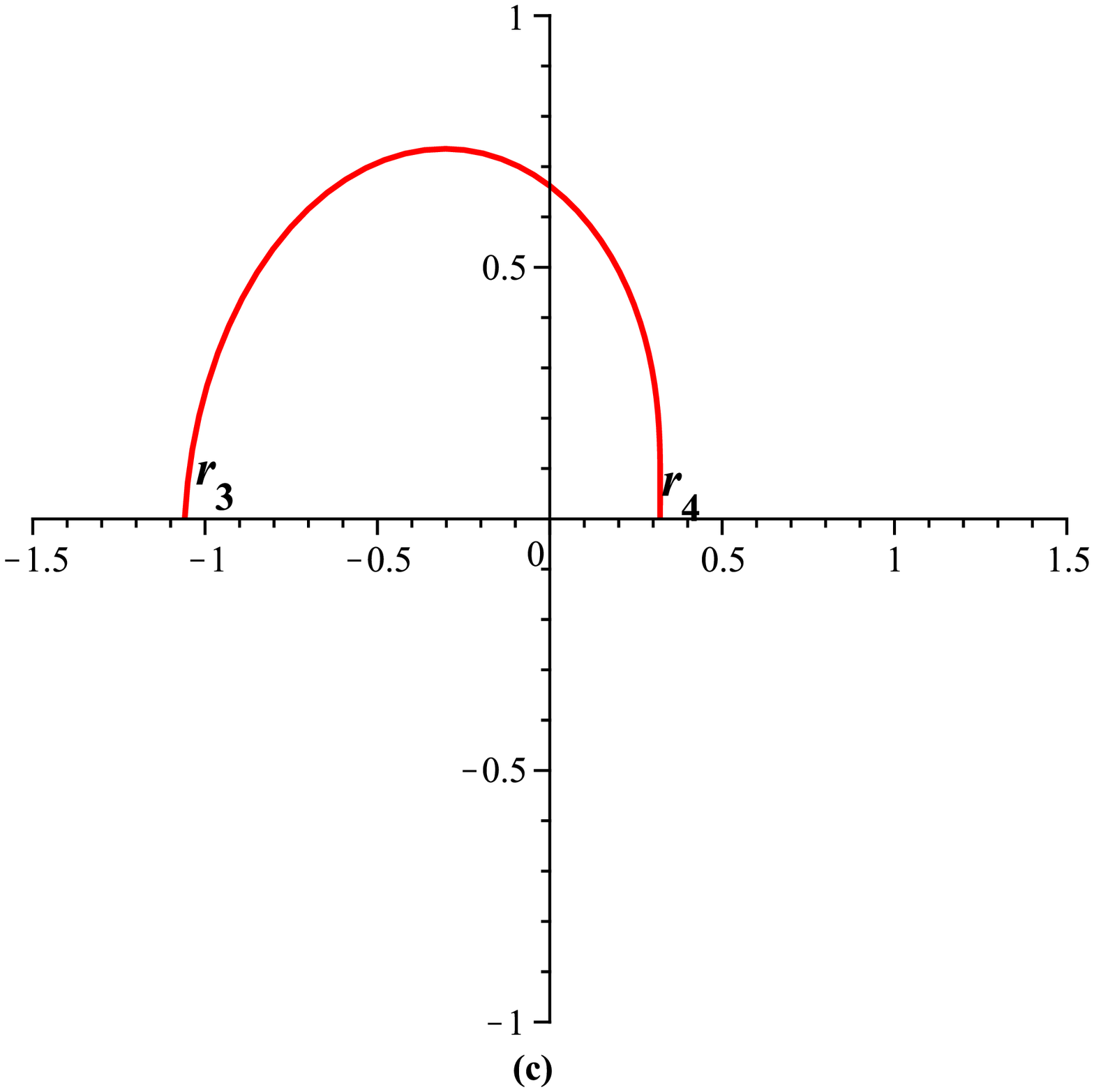}}
\caption{
(a) $f$(u) represents the shaded allowed region for the motion of photons with four real roots at points $\bf{u_1}$, $\bf{u_2}$, $\bf{u_3}$ and $\bf{u_4}$, with $D$ = 3.5, $Q$ = 0.4 and $m$ = 0.5;(b) Solid line represents the {\it fly-by orbit} for the null geodesics approaching the BH from infinity with a turning point at $\bf{r_2}$ and again flying back to infinity; (c) Solid line represents a {\it bound orbit} from $\bf{r_3}$ to $\bf{r_4}$. In fig.(a) $u$ is used as a coordinate while $r$ is used as a coordinate in figs. (b) and (c).}
\protect\label{f8}
\end{figure}

\newpage
\noindent One may observe from fig.(\ref{f5}) that the radius of {\it unstable circular orbit} is smallest for MCBH and $r^M_{c}<r^S_{c}<r^E_{c}$.
Hence, in comparison to the SBH, the gravitational field of MCBH is more attractive in nature while the gravitational field for ECBH is rather repulsive.
This particular nature of the gravitational field around the respective BH spacetime can also be confirmed from the nature of respective effective potentials as depicted in fig.(\ref{com_v}a).
\\
\vspace{2mm}\\
\underline{\bf{The Time Period:}}\\
The time period for {\it unstable circular orbit} can be calculated for proper time as well as
coordinate time with $\phi$ = 2 $\pi${\cite{fern12}}.
For MCBH, on integrating eq.(\ref{eqn:first_intg_M}) for one complete time period,\\
\begin{equation}
T_{\tau} = \frac{2 \pi r_c^2}{L} ,
\end{equation}
represents the time period in proper time while dividing eq.(\ref{eqn:first_intg_M}) with eq.(\ref{eqn:first_int_t}) and integrating for one complete time period,
\begin{equation}
T_{t} = 2 \pi \frac{r_c \sqrt{m r_c - Q^2}}{\sqrt{m (r_c - 2 m)}} ,
\end{equation}
represents the time period in coordinate time.
The corresponding expressions for  ECBH are,
\begin{equation}
T_{\tau,E}=\frac{2\pi r_c^2}{L} ,
\end{equation}
and
\begin{equation}
T_{t,E}=2\pi (r_c+2 m sinh^2 \alpha)\sqrt{\frac{r_c}{r_c-2m}} .
\end{equation}
In the limits $Q \rightarrow$ 0 and $\alpha\rightarrow$ 0, the expressions for orbital time periods in proper time ( $T_{\tau}$ and $T_{\tau,E}$) and coordinate time ( $T_{t}$ and $T_{t,E}$ ) reduce exactly to that for the SBH i.e, $T_{\tau,S}=\frac{2\pi (3 m)^2}{L}$ and $T_{t,S}= 3 \sqrt{3} m$.\\
\begin{figure}[h!]
\centerline{\includegraphics[width=6.5cm,height=6cm]{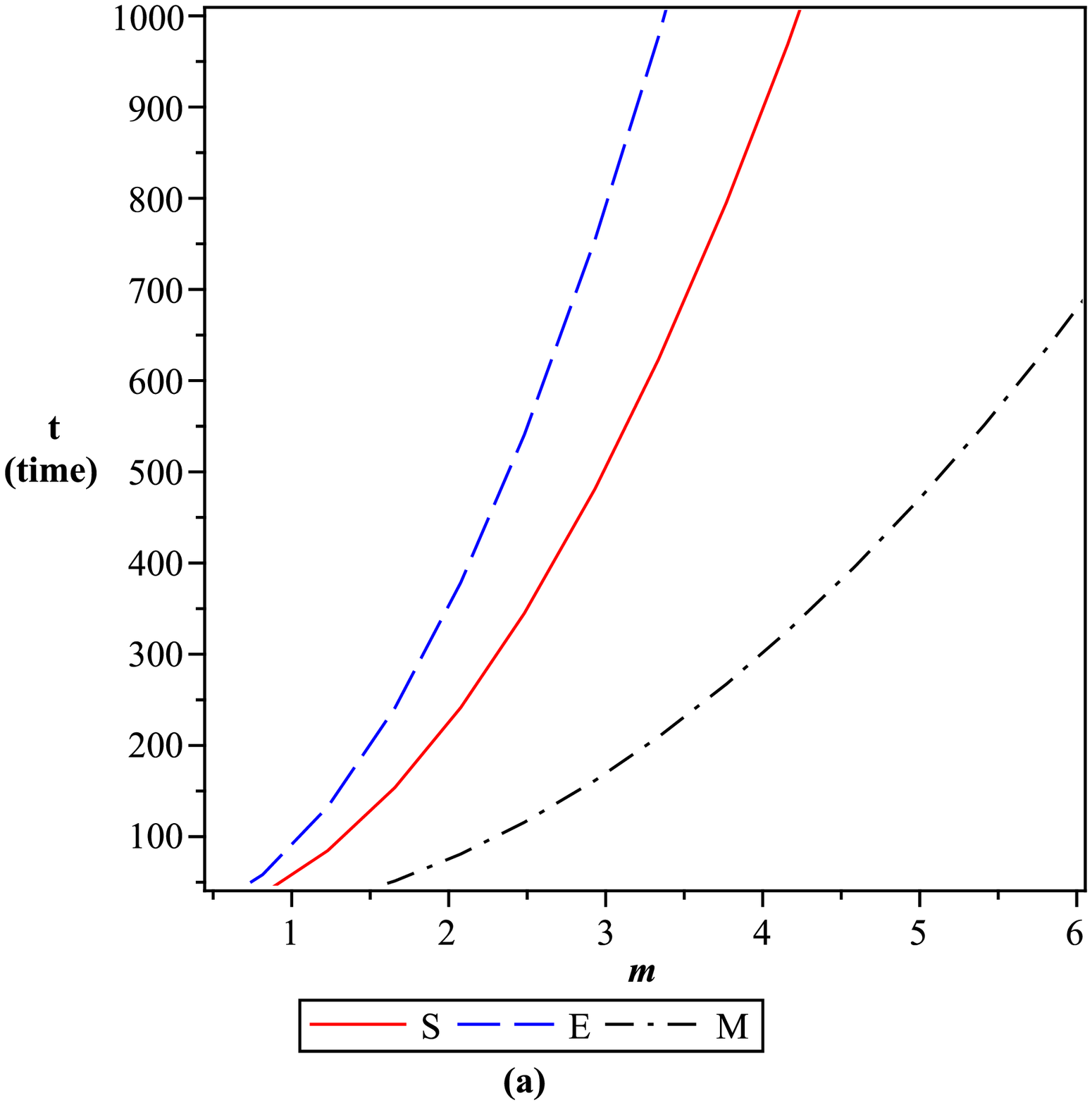}
\includegraphics[width=7cm,height=6cm]{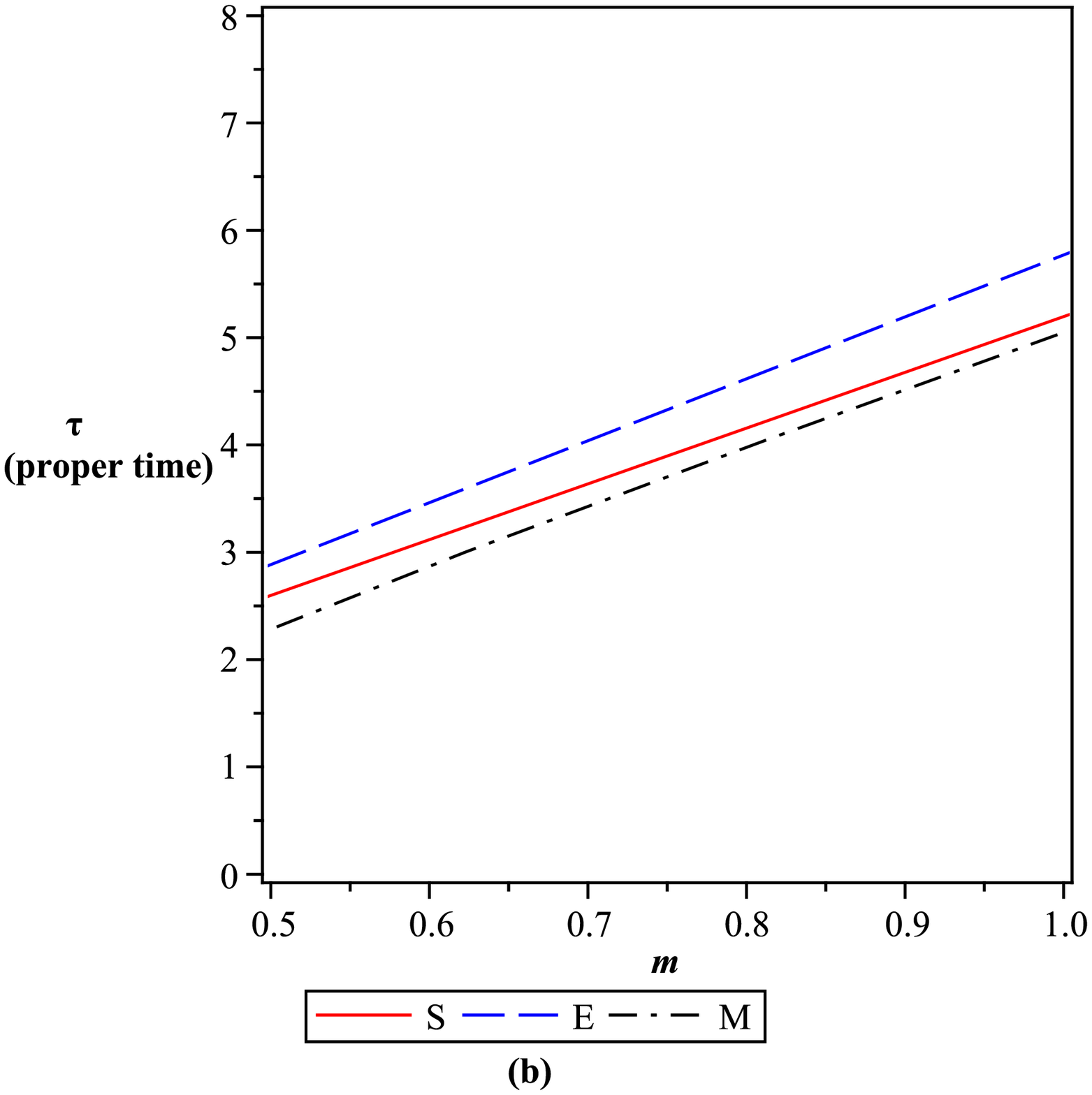}}
\caption{Variation of time periods for; (a)coordinate time and (b)proper time with mass $m$ where SBH, ECBH and MCBH, represented by the curves $S$, $E$ and $M$ respectively; with $Q$ = $\alpha$ = 0.4.}
\protect\label{f6}
\end{figure}\\
\noindent Fig.(\ref{f6}) depicting the time periods for all three BHs clearly shows that the time periods for MCBH are smaller than SBH while the respective values are larger for the case of ECBH.
This result is in agreement with the results obtained from the comparison of radius of {\it unstable circular orbits} and the nature of effective potentials for the BHs.
\\
\vspace{2mm}\\
\underline{\textbf{Frequency Shift:}}\\
Here, as one of the optical phenomenon, we calculate the combined gravitational and Doppler frequency shift for MCBH in the equatorial plane.
The Keplerian angular frequency of the circular geodesics relative to distant observers is given by,
\begin{equation}
\Omega = \frac{d \phi}{d t}.
\label{freq}
\end{equation}
Using eq.(\ref{eqn:first_intg_M}) in eq.(\ref{freq}), we obtain,
\begin{equation}
\Omega = \frac{L}{E}\frac{(1-\frac{2 m}{r})}{r^2\left(1 - \frac{Q^2}{m r}\right)}.
\label{freq_M1}
\end{equation}
From the general expression for the frequency shift in equatorial plane \cite{stu14}, the frequency shift for MCBH is given by,
\begin{equation}
1 + z =\frac{\sqrt{(1 - \frac{2 m}{r})}}{\sqrt{(1 - \frac{Q^2}{m r})\left[1 -\frac{D^2}{r^2}\left(\frac{1 - \frac{2 m}{r}}{1 - \frac{Q^2}{m r}}\right)\right]}}.
\label{freq_shft_f_M}
\end{equation}
The frequency shift depends on the charge parameter of the BH.
As the frequency shift increases, the observed frequency of the photon decreases which gives an equivalent increase in the corresponding wavelength of the photon. Hence for positive frequency shifts, the photon is redshifted.
Again stronger is the gravitational field of the gravitating source, larger will be the energy loss by the incoming photon and larger will be the observed redshhift.\\
\begin{figure}[h!]
\centerline{\includegraphics[width=6cm,height=5cm]{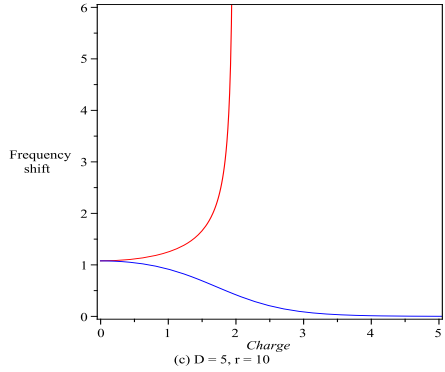}}
\caption{Variation of frequency shift with charge parameter for MCBH and ECBH.}
\protect\label{f7}
\end{figure}
\\
Fig.(\ref{f7}) depicts the comparative redshifts in the equatorial plane for MCBH and ECBH at some particular radial distance $r$.
The redshift for photons around the MCBH is increasing with increase in the magnetic charge while it decreases for the case of ECBH.
Hence at fixed radial distance $r$, the gravitational field of MCBH strengthens if magnetic charge is increased while the the gravitational field of ECBH weakens if electric charge is increased.\\
\vspace{2mm}\\
\underline{\textbf{The Cone of Avoidance:}}\\
If now a point source at a given distance $r$ of the centre emits light isotropically into all directions, a part of the light will be captured by the BH, while another part can escape.
It is clear that the critical orbits are at the limit of infall or escape. These orbits define a cone with half opening angle $\Psi$. The \textit{cone of avoidance} is defined by null rays being its generator, described by the eq.(\ref{eqn:sol_orbit_M}), passing through that point \cite{cha83,nor05}.
If $\Psi$ denotes the half angle of the cone then,
\begin{equation}
\cot\Psi = \frac{1}{r}\frac{d\tilde{r}}{d\phi} ,
\label{eq:cot-psi}
\end{equation}
where $\tilde{r}$ is the proper length along the generators of the cone,
\begin{equation}
d\tilde{r}=\frac{1}{\sqrt{(1-\frac{Q^2}{mr})(1-\frac{2m}{r})}} dr .
\label{eq:tilde-r}
\end{equation}
Now, using eq.(\ref{eqn:u-phi_M}) and eq.(\ref{eq:tilde-r}) in eq.(\ref{eq:cot-psi}),
\begin{equation}
\tan\Psi=\frac{u \sqrt{1-2 m u}}{\sqrt{2m(u-u_1)(u-u_c)^2}} .
\end{equation}
When $u\rightarrow u_c (r\rightarrow r_c), \Psi\rightarrow \frac{\pi}{2}$ and when $u\rightarrow 0 (r\rightarrow \infty), \Psi\approx (\frac{1}{r}) \sqrt{\frac{-1}{2 m u_1 u_c^2}}$. In the limit $Q\rightarrow 0$, $\Psi$ approaches to the corresponding values for the SBH. One can immediately see that at large distances the cone is narrow, and most of the light escapes.
For $r$ = $r_c$, exactly half of the light is being captured by the BH. For $r$ = 2$m$, $\Psi$ = 0 and therefore the horizon is the final frontier from where photon cannot escape and at larger distances $\Psi$ depends on the corresponding charge of the BH.\\
\vspace{3mm}\\
\textbf{Case (b) : Photon orbits with $D > D_c$}\\
Now, let us choose the value of impact parameter $D$ = 3.5, which is larger than the impact parameter for the case of unstable circular orbit for all three BHs discussed.
As shown in Fig.(\ref{f8}a), $f$(u) has three positive real roots for $D > D_c$.
Hence to obtain the analytic solution of orbit, $f$(u) can be expressed as $f(u)=-Q^2 (u-u_1)(u-u_2)(u-u_3)(u-u_4)$ in eq.(\ref{eqn:u-phi_M}), which on integration yields,
\begin{equation}
\phi+\phi_{{0}}=-{\frac {P \sqrt {2}}{\sqrt {-{Q}^{2} \left( u_1-u_3
 \right)  \left( u_2-u_4 \right) }}} ,
\label{eqn:sol_orbit_M2}
\end{equation}
where,
\begin{equation}
P={ Elliptic\hskip2mmF} \left( \sqrt {{\frac { \left( u_1-u_3 \right)  \left( -u+u_4
 \right) }{ \left( u_1-u_4 \right)  \left( -u+u_3 \right) }}},\sqrt {{\frac
{ \left( -u_3+u_2 \right)  \left( u_1-u_4 \right) }{ \left( u_2-u_4 \right)
 \left( u_1-u_3 \right) }}} \right).
\end{equation}
\begin{figure}[h!]
\centerline{\includegraphics[width=5cm,height=7cm]{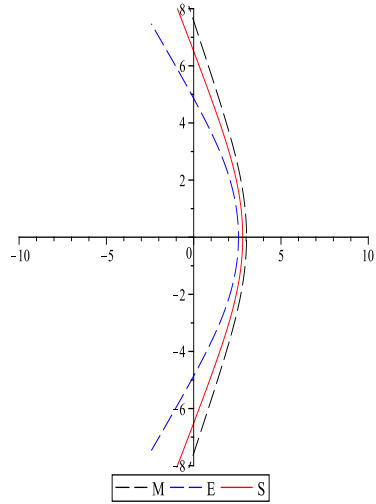}}
\caption{Comparative view of {\it fly-by orbits} for null geodesics with $Q$ = $\alpha$ = 0.4;  where curves $S$, $E$ and $M$ represent the SBH, ECBH and MCBH respectively.}
\label{f9}
\end{figure}
\newpage
\noindent The structure of {\it fly-by orbits} (or scattering  orbits) is qualitatively similar for all the three BHs discussed.
One can observe from fig.(\ref{f9}) that the turning point for the corresponding fly-by orbit is farthest for MCBH and ${r_E}<{r_S}<{r_M}$.\\
\vspace{3mm}\\
\textbf{Case (c) : Photon orbits with $D < D_c$}\\
To analyse the orbit structure for $D < D_c$, the impact parameter is now fixed at $D = 1$. Integrating eq.(\ref{eqn:u-phi_M}), we obtain the same solution as in eqn.(\ref{eqn:sol_orbit_M2}) but now $u_2$ and $u_3$ are imaginary.
\begin{figure}[h]
\centerline{\includegraphics[width=6cm,height=6cm]{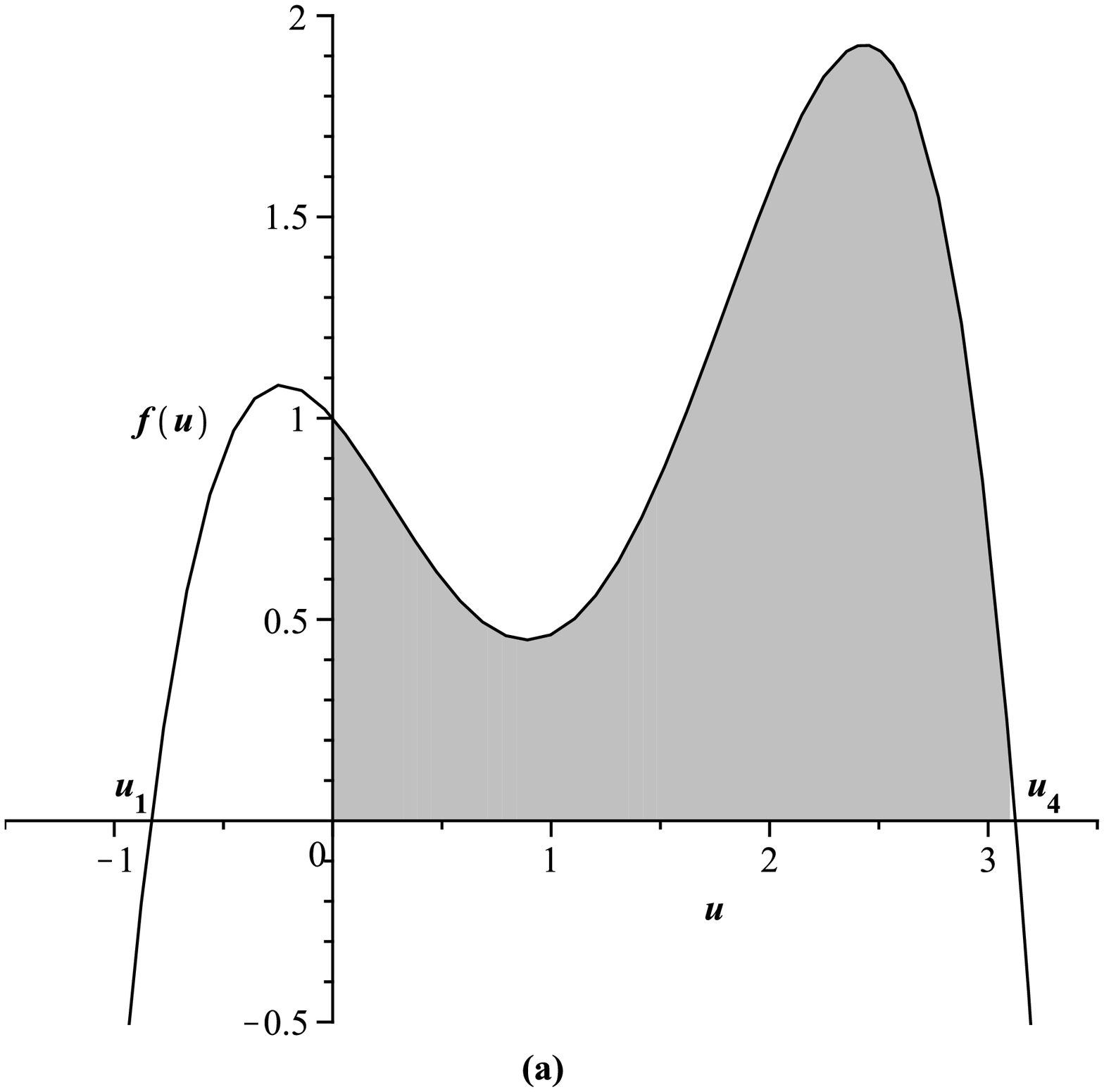}
\includegraphics[width=6cm,height=6cm]{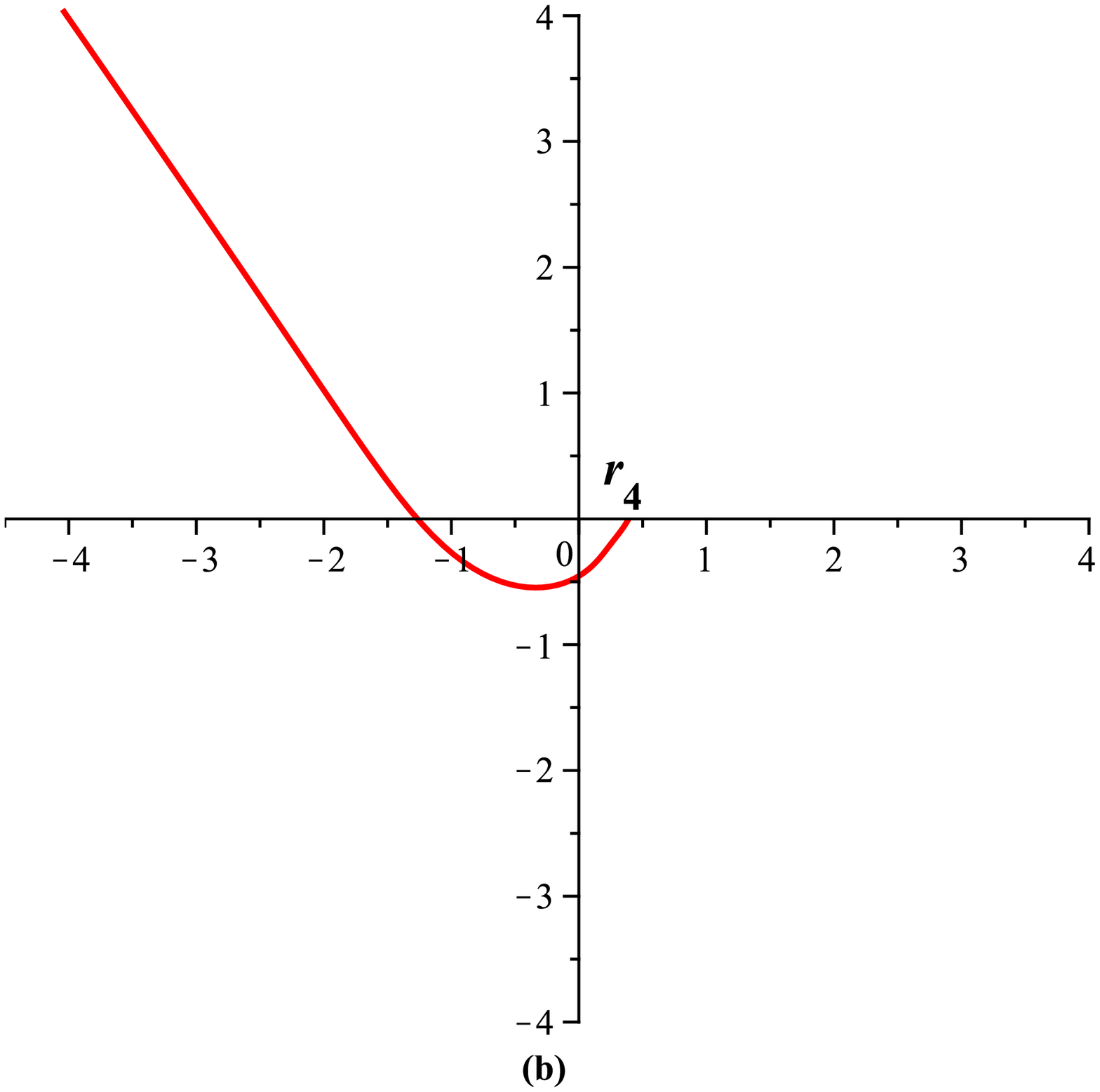}}
\caption{
(a) $f$(u) represents the shaded allowed region for the motion of photons with two real roots at points $\bf{u_1}$ and $\bf{u_4}$, with $D$ = 1, $Q$ = 0.4 and $m$ = 0.5;(b) Solid line represents the {\it terminating escape orbit} for the null geodesics approaching the BH from infinity to $\bf{r_4}$ while $\bf{r_4}$ lies inside the event horizon of the BH. In fig.(a) $u$ is used as a coordinate while $r$ is used as a coordinate in fig.(b).}
\protect\label{f10}
\end{figure}
\begin{figure}[h!]
\centerline{\includegraphics[width=5cm,height=6cm]{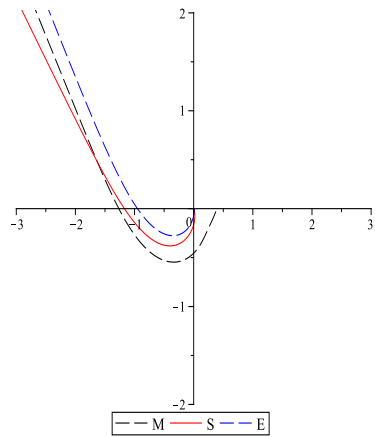}}
\caption{Comparative view of {\it terminating escape orbits} with $Q$ = $\alpha$ = 0.4; where $S$, $E$ and $M$ represent the SBH, ECBH and MCBH respectively.}
\protect\label{f11}
\end{figure}
\newpage
\noindent One can observe from fig.(\ref{f11}) that ${r_E}<{r_S}<{r_M}$ similar to the case of $D>D_c$ and the {\it terminating escape orbits}, end up at the corresponding curvature singularity of the BH spacetime.
Hence for MCBH the orbit terminates at point $r_4$ while for SBH and ECBH the terminating point is at $r$ = 0.

\section{Discussion and conclusion}
\noindent In this article, we have investigated the geodesic motion for null geodesics in the background of a MCBH arising in string theory and some of the important results obtained are summarised below.
\begin{itemize}
  \item The nature of effective potential for MCBH is qualitatively similar to that of SBH and ECBH. But, as the impact parameter for the incoming photon changes w.r.t. the corresponding value for the {\it unstable circular orbit}, the attractive nature of the effective potential changes accordingly. This change in the nature of effective potential manifests in the corresponding orbit structure also.

  \item The different types of orbits (such as unstable circular orbit, fly-by orbit, terminating escape orbit, bound orbit) are present for the incoming photons, corresponding to the value for the impact parameter. No stable circular orbits are present for the null geodesics.

  \item The gravitational field of MCBH is found to be more attractive in nature than SBH and ECBH, for the incoming photons with $D$ = $D_c$. This effect is manifested in the results of the radius of unstable circular orbit, time period for proper time as well as coordinate time and frequency shift for null geodesics.

\item The attractive nature of gravitational field of MCBH strengthens as the magnetic charge increases while for ECBH it weakens as electric charge increases.
Interestingly as the impact parameter differs from $D_c$, the gravitational field of MCBH becomes more repulsive than SBH and ECBH.

  \item Cone of avoidance for large distances is found to depend on the respective charges of MCBH and ECBH. At larger distances, the cone is narrow and therefore most of the photons can escape from there.
\enditemize
We believe that the result obtained herewith would be useful in the study of gravitational lensing around the charged BHs in the string theory and we intend to report on this issue in near future.
\end{itemize}
\section*{Acknowledgments}
\noindent The authors would like to thank Department of Science and Technology, New Delhi for the financial support through grant no. SR/FTP/PS-31/2009. The authors are also thankful to Buddhi V
Tripathi for the useful discussions and comments.

\end{document}